\documentclass[twocolumn,
               showpacs,
               preprintnumbers,
               nofootinbib,
               prd,
               superscriptaddress,
               10pt,
               notitlepage,
               aps]{revtex4-1}

\usepackage[T1]{fontenc}

\usepackage{amssymb}
\usepackage{amsmath}
\usepackage{amsfonts}
\usepackage{amsbsy}
\usepackage{dsfont}
\usepackage{mathrsfs}
\usepackage{bm}
\usepackage[inline]{enumitem}
\usepackage{float}

\usepackage[linktocpage,breaklinks]{hyperref}
\usepackage[usenames,dvipsnames]{color}

\usepackage{epsfig}
\usepackage{graphicx}
\usepackage{epstopdf}

\usepackage[linktocpage]{hyperref}
\usepackage[usenames,dvipsnames]{color}
\usepackage{empheq}
\usepackage{placeins}

\hypersetup{colorlinks=true, citecolor=NavyBlue,
            linkcolor=NavyBlue, urlcolor=RedViolet}

\newcommand{\dd}{{\rm d}}
\newcommand{\io}{\iota_{\rm o}}
\newcommand{\ts}{\theta_{\rm s}}
\newcommand{\ag}{AlGendy~\&~Morsink\,}
\newcommand{\ms}{Morsink~et~al.\,}

\begin{document}

\title{The surface of rapidly-rotating neutron stars: \\ implications
       to neutron star parameter estimation}

\author{Hector O. Silva}
\affiliation{Illinois Center for Advanced Studies of the Universe \&
Department of Physics, University of Illinois at Urbana-Champaign,
Urbana, Illinois 61801, USA}
\affiliation{eXtreme Gravity Institute, Department of Physics, \\
Montana State University, Bozeman, Montana 59717 USA}
\affiliation{Max Planck Institute for Gravitational Physics (Albert Einstein Institute), \\
Am M\"uhlenberg 1, Potsdam D-14476, Germany}

\author{George Pappas}
\affiliation{Dipartimento di Fisica ``Sapienza''
Universit\`a di Roma \& Sezione INFN Roma1, \\
Piazzale Aldo Moro 5, 00185, Roma, Italy}
\affiliation{Department of Physics, Aristotle University of Thessaloniki,
Thessaloniki 54124, Greece}

\author{Nicol\'as Yunes}
\affiliation{Illinois Center for Advanced Studies of the Universe \&
Department of Physics, University of Illinois at Urbana-Champaign,
Urbana, Illinois 61801, USA}
\affiliation{eXtreme Gravity Institute, Department of Physics,
Montana State University, Bozeman, Montana 59717, USA}

\author{Kent Yagi}
\affiliation{Department of Physics, University of Virginia,
Charlottesville, Virginia 22904, USA}

\begin{abstract}
The \emph{Neutron star Interior Composition Explorer} (NICER) is currently
observing the x-ray pulse profiles emitted by hot spots on the surface of
rotating neutron stars allowing for an inference of their radii with
unprecedented precision.
A critical ingredient in the pulse profile model is an analytical formula
for the oblate shape of the star.
These formulas require a fitting over a large ensemble of neutron star
solutions, which cover a wide set of equations of state, stellar compactnesses
and rotational frequencies.
However, this procedure introduces a source of systematic error, as
%%%%
\begin{enumerate*}[label={(\roman*)}]
\item the fits do not describe perfectly the surface of \emph{all} stars in the
ensemble and
\item neutron stars are described by a single equation of state, whose influence on
the surface shape is averaged out during the fitting procedure.
\end{enumerate*}
%%%%
Here we perform a first study of this systematic error, finding evidence that
it is subdominant relative to the statistical error in the radius inference
by NICER.
We also find evidence that the formula currently used by NICER
can be used in the inference of the radii of rapidly
rotating stars, outside of the formula's domain of validity.
Moreover, we employ an accurate enthalpy-based method to locate the surface
of numerical solutions of rapidly rotating neutron stars and a new
highly accurate formula to describe their surfaces.
These results can be used in applications that require an accurate
description of oblate surfaces of rapidly rotating neutron stars.
\end{abstract}

\date{\today}
\maketitle

\section{Introduction}
\label{sec:intro}

One of the outstanding open problems in nuclear astrophysics is the
determination of the properties of cold, nuclear matter above nuclear
saturation density.
The physics of nuclear matter in this scenario is encapsulated in the so-called
(barotropic) equation of state: a relation between the pressure and energy
density of matter, assumed to be described by a perfect fluid.
In this context, neutron stars provide a natural laboratory to explore the
equation of state.
The precise inference of neutron star properties, such as
the masses $M$ and (equatorial) radii $R_{\rm eq}$, is expected to
reveal the equation of state~\cite{Lattimer:2006xb,Baym:2017whm}.
While the former has been measured with exquisite precision through
measurements of the orbital parameters in double pulsar system with radio
astronomy, the latter remains elusive, with current
inferences having large systematic errors~\cite{Ozel:2012wu,Miller:2016pom,Degenaar:2018lle}.

The \emph{Neutron star Interior Composition Explorer} (NICER) is currently
observing the x-ray emission from hot spots
on the surface of neutron
stars~\cite{Gendreau2012:SPIE,Arzoumanian2014:SPIE,Gendreau2017:NatAst}.
This x-ray flux is seen as a pulsation in a detector and its shape (i.e. the
profile) carries information about the surface properties of the star and
the spacetime surrounding it~\cite{Watts:2016uzu,Watts:2019lbs}.
Combined, this information allows for the simultaneous inference of both the
mass $M$ and the equatorial radius $R_{\rm eq}$ at the 5\%--10\% level.
The mission's promise was recently realized with the announcement of the measurement of the mass
and (equatorial) radius of the \emph{isolated} millisecond pulsar the
J0030+0451~\cite{Riley:2019yda,Miller:2019cac}, demonstrating the usefulness
of time and energy-resolved x-ray observations to infer neutron star properties.
Moreover, additional properties (such as the star's moment of inertia) can be
inferred using quasi-equation-of-state independent relations~\cite{Silva:2020acr}.
Further inferences obtained from the observation of three other pulsars PSRs~J0437--4715,
J1231--1411, and J2124--3358 are expected to be released in the near
future~\cite{Bogdanov:2019ixe}.
These electromagnetic observations combined with gravitational-wave inferences
on the tidal deformability from neutron star binaries will improve
considerably our understanding of the neutron star equations of state (see
e.g.~\cite{Raithel:2019uzi,Raaijmakers:2019qny,Raaijmakers:2019dks,Jiang:2019rcw,Zimmerman:2020eho,Dietrich:2020efo,Chatziioannou:2020pqz,Essick:2020flb}).

In principle, the pulse profile can be calculated by performing ray-tracing
from the hot spot(s) to the observer in numerically constructed neutron star spacetime
models~\cite{Cadeau:2004gm,Cadeau:2006dc}.
In practice, the large multidimensional parameter space of the problem makes it
computationally prohibitive to use ray-tracing for parameter inference using
Bayesian methods.
This obstacle calls for a pulse profile model that is computationally efficient to calculate, yet
captures the salient features of a full ray-tracing calculation.
In the canonical pulse profile model used in the literature, photons are
emitted from an oblate surface and assumed to propagate in an ambient
Schwarzschild background~\cite{Cadeau:2006dc,Morsink:2007tv}.
Previous works~\cite{Cadeau:2006dc,Bogdanov:2019qjb} have shown that this
``Oblate+Schwarzschild'' (O+S) approximation provides all the necessary ingredients to
capture, with good precision, the results of ray-tracing in numerically
generated neutron star spacetimes\footnote{An earlier subset of this model in which the
star is spherical is known as the ``Schwarzschild+Doppler'' approximation~\cite{Miller:1997if}.
(See also~\cite{Poutanen:2003yd,Poutanen:2006hw}).}.

The O+S model takes as an ingredient an analytical formula to describe the
rotation-induced oblateness of the star~\cite{Morsink:2007tv}.
The use of such ``shape formulas'' by-pass the process of calculating numerically rapidly
rotating neutron star models~\cite{Paschalidis:2016vmz}, which is also
computationally expensive in itself.
Such formulas have been suggested in the
literature~\cite{Morsink:2007tv,AlGendy:2014eua} and they share the feature of
being obtained by fitting an analytically prescribed ``shape function'' to a
large ensemble of rotating neutron star models, which covers a large sample of
equations of state and spin frequencies.
This fitting process introduces a systematic error when estimating
e.g. the star's radius because
%
%%%%
\begin{enumerate*}[label={(\roman*)}]
\item the fits do not describe perfectly the surface of \emph{all} stars in the
ensemble and
\item neutron stars are, in reality, described by a single equation of state, whose influence on
the surface shape is averaged out during the fitting procedure.
\end{enumerate*}
%%%%
%
As the shape of a star being observed is determined by its rotation
frequency and its underlying equation of state,
the radius inference is, in principle, affected by the ensemble used to find the fit.

Having identified that this may be a source of systematic error, it is natural to ask if
it has an immediate impact on NICER today, or in the future.
Here we perform a first study on this issue.
We first numerically construct rotating neutron star solutions, valid to all
orders in rotation, and compare their surface to the different fits used in the
literature.
We then create a new fitting function that is better suited at recovering the
surface of rapidly rotating neutron stars.
With these fitting functions at hand, we then study through a simplified
Bayesian analysis whether the use of fitting functions introduces systematic
errors in the parameters extracted.
We find evidence that this
systematic error is subdominant relative to the statistical error in the
radius inference by NICER.
We also find evidence that the formula currently used by NICER
can be used in the inference of the radii of rapidly
rotating stars, outside of the formula's domain of validity.

In the remainder of this paper we present how we arrived at these conclusions.
In Sec.~\ref{sec:surface} we present the neutron star models we use, how they
are computed and present a method to accurately locate their surfaces.
We also review how the fitting formulas for neutron star surfaces are obtained
and introduce a new formula that describes accurately the surface of
rapidly rotating neutron stars.
In Sec.~\ref{sec:lightcurves} we analyze in detail the impact of the different
fitting formulas on the resulting pulse profile and their impact on the
inference of the equatorial radius.
In Sec.~\ref{sec:conclusions} we summarize our conclusions and discuss possible
extension of this work.
Unless stated otherwise, we work in geometric units with $c = 1 = G$.

\section{The surface of rotating neutron stars}
\label{sec:surface}

\subsection{Rapidly rotating neutron stars}
\label{sec:rns}

We start by calculating a large catalog of rapidly, rotating neutron star
solutions using the RNS (``rotating neutrons stars'') code developed by
Stergioulas and Friedmann~\cite{Stergioulas:1994ea}.
The code obtains equilibrium neutron star solutions by solving Einstein's
equations in the presence of a perfect fluid using the
Komatsu-Eriguchi-Hachisu scheme~\cite{Komatsu:1989zz,Komatsu:1989za} and
improving upon the modifications introduced by Cook, Shapiro and
Teukolsky~\cite{Cook:1993qr,Cook:1993qj}.
All these methods use the line element of a stationary and axisymmetric spacetime, which, in quasi-isotropic coordinates, is given by:
\begin{align}
\dd s^2 &=
- \, e^{\gamma + \rho}\, \dd t^2
+ e^{2\alpha}\left( \dd r^2 + r^2 \, \dd \theta^2 \right)
\nonumber \\
&\quad + r^2 e^{\gamma - \rho} \sin^2\theta\, (\dd \phi - \omega\, \dd t)^2\,,
\label{eq:line_element}
\end{align}
where $\alpha$, $\gamma$, $\rho$ and $\omega$ are
functions of the coordinates $r$ and $\theta$ only.
Given a rotation law (we assume uniform rotation) and an equation of state, the RNS code can
obtain equilibrium solutions once a central energy density
$\varepsilon_{\rm c}$ and a ratio $r_{\rm pol}/r_{\rm eq}$
(between the polar and the equatorial coordinate radii) have been specified.

\begin{figure*}[t]
\includegraphics[width=\columnwidth]{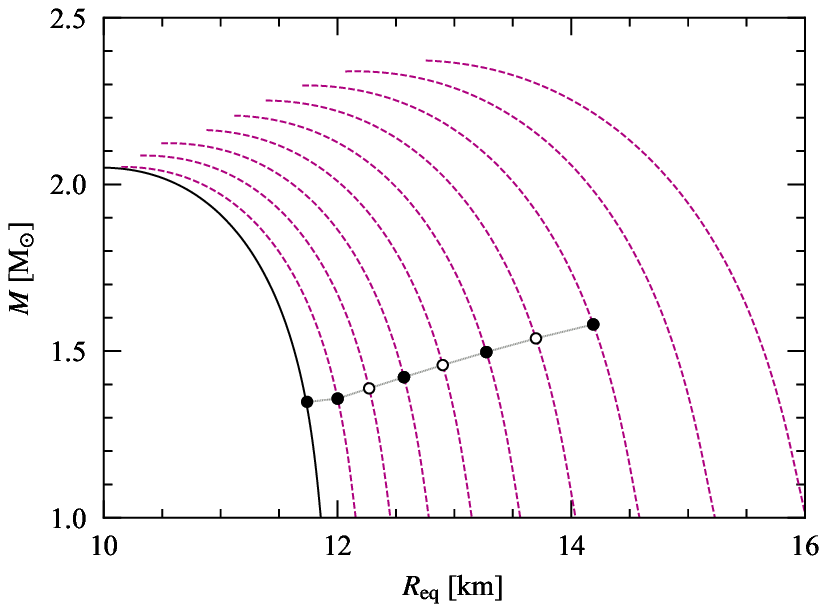}
\includegraphics[width=\columnwidth]{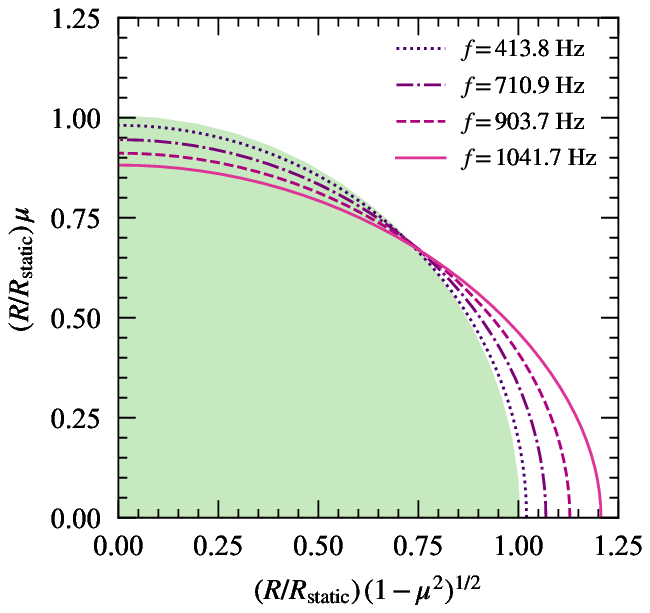}
\caption{Left: families of rotating neutron star solutions parametrized by their central
energy density, using the SLy4 equation of state and evenly spaced
in polar-to-equatorial coordinate radius ratio $r_{\rm pol}/r_{\rm eq}$,
from static configuration (solid, leftmost curve) to their
Kepler limit (dashed, rightmost curve).
As the rotation frequency increases, the stars with fixed central
energy density move toward larger equatorial radii and masses.
This is illustrated by the dots, which correspond to a fixed central energy
density, with parameters presented in
Table~\ref{tab:ref_stars}.
Right: illustration of the deformation of a neutron star caused
by rotation.
The curves show the surface of the stars marked by a black circle in the left panel.
The axes correspond to $(R / R_{\rm static}) \, (1-\mu^2)^{1/2}$ and $(R / R_{\rm static}) \, \mu$.
The shaded region represents a static, spherically
symmetric star using the SLy4 equation of state and central total
energy density $\varepsilon_{\rm c}/c^2 = 9.4769 \times 10^{14}$ g/cm$^3$,
with a radius of $R_{\rm static} = 11.76$ km and
a mass of $M = 1.35$~M$_{\odot}$.
When the rotation frequency $f$ increases, the star flattens at the poles,
while bulging out in the equator, as shown by the various lines.
The ratio $\mathfrak{r}$ between the polar $R(1)$ to the equatorial radius
$R(0)$ decreases from unity (for the static model) to $0.72$ at
$f = 1041.75$ Hz.
}
\label{fig:mass_radius}
\end{figure*}

Once a neutron star solution has been obtained, we can determine the star's
coordinate surface $r_{\rm s}(\theta)$ by the loci where the pressure vanishes. Then,
the (circumferential) radius of the star is determined as a function of the cosine
of the colatitude $\theta$ as,
\begin{equation}
R(\mu) = r_{\rm s} \, e^{(\gamma_{\rm s} - \rho_{\rm s})/2}\,,
\label{eq:circ_radius}
\end{equation}
where we defined $\mu \equiv \cos\theta$, $\gamma_{\rm s} \equiv \gamma(r_{\rm s})$
and $\rho_{\rm s} \equiv \rho(r_{\rm s})$.
Based on this definition of the surface, we also define for later use the ratio
$\mathfrak{r}$
\begin{equation}
\mathfrak{r} \equiv {R_{\rm pol}} / {R_{\rm eq}}\,,
\label{eq:ratio_poleq}
\end{equation}
between polar radius [$R_{\rm pol} \equiv R(\mu=1)$] and
equatorial radius [$R_{\rm eq} \equiv R(\mu=0)$].
We further define the eccentricity of the star as
\begin{equation}
e \equiv \sqrt{1 - \mathfrak{r}^2}\,,
\quad\textrm{(eccentricity)}.
\end{equation}

To remain agnostic regarding the underlying matter description of neutron star
interiors, we consider a set of equations of state that covers a wide variety of
predicted neutron star masses and radii.
The equations of state we use, in increasing order of stiffness (i.e.~largest
maximum mass supported) are:
FPS~\cite{Baym:1971pw,Lorenz:1992zz},
SLy4~\cite{Douchin:2001sv},
AU~\cite{Wiringa:1988tp,Negele:1971vb},
UU~\cite{Wiringa:1988tp,Negele:1971vb},
APR~\cite{Akmal:1998cf} and
L~\cite{Pandharipande:1976zz}.
Most of these equations of state are consistent with the recent gravitational
wave observations of a binary neutron star coalescence by the LIGO Scientific
Collaboration~\cite{Abbott:2018exr}, with the exception of FPS and L, which are
not stiff enough and too stiff respectively, but we include them here
nonetheless for completeness.
For each equation of state, we calculate 198 equilibrium configurations
parametrized by the central energy density $\varepsilon_{\rm c}$ and evenly
spaced in the polar-to-equatorial coordinate radii ratio $r_{\rm pol}/r_{\rm
eq}$, from slowly rotating models up to the Kepler limit.
In total, our catalog consists of 1188 stars.

To illustrate the impact of rotation on the properties of neutron stars, we show in
the left panel of Fig.~\ref{fig:mass_radius} the mass-(equatorial) radius relation for a
family of solutions obtained using the SLy4 equation of state and various rotation rates.
The solid line represents the nonrotating family of solutions, obtained by
integrating the TOV (Tolman-Oppenheimer-Volkoff)
equations~\cite{Tolman:1939jz,Oppenheimer:1939ne} for a range of
central energy densities~$\varepsilon_{\rm c}$.
The dashed lines represent families of solutions with increasing
$r_{\rm pol} / r_{\rm eq}$ ratios, which is equivalent to an increase of the
rotational frequency $f$.
We see that the mass-(equatorial) radius relations shifts toward
larger radii (due to the ``bulging'' out of the star's equator) and larger masses
(due to the contribution of rotational energy to the star's gravitational
mass and more support to baryons).

This behavior becomes more evident by tracking stars
with constant~$\varepsilon_{\rm c}$ as we increase $r_{\rm pol} / r_{\rm eq}$.
As an example, in the left panel of Fig.~\ref{fig:mass_radius} we mark with circles the solutions with $\varepsilon_{\rm c} =
9.4769 \times 10^{14}$ g/cm$^{3}$, which show the trend described above.
This particular sequence of stars covers rotation frequencies $f$ between
approximately 400 and 1050~Hz, and will later serve as benchmark in our work.
A variety of their properties are summarized in Table~\ref{tab:ref_stars}, and
their surfaces (obtained by a procedure described in
Sec.~\ref{sec:how_to_fit}) are shown in the right panel of
Fig.~\ref{fig:mass_radius}.

\begin{table}[t]
\begin{tabular}{c c c c c c c c c}
\hline
\hline
Model & $M$ & $R_{\rm eq}$ & $\mathfrak{r}$  & $j$ & $q$ & $f$ & $\sigma$ & $\kappa$ \\
      & ($M_{\odot}$) & (km) & & & & (Hz)& \\
\hline
1 & 1.377 & 12.00 & 0.956 & 0.215 & 5.225 & 413.8  & 0.064 & 0.169 \\ \hline
2 & 1.408 & 12.27 & 0.912 & 0.307 & 4.940 & 583.4  & 0.133 & 0.169 \\ \hline
3 & 1.442 & 12.57 & 0.868 & 0.381 & 4.685 & 710.9  & 0.207 & 0.169 \\ \hline
4 & 1.479 & 12.90 & 0.824 & 0.444 & 4.445 & 815.4  & 0.287 & 0.169 \\ \hline
5 & 1.518 & 13.27 & 0.780 & 0.501 & 4.222 & 903.7  & 0.374 & 0.169 \\ \hline
6 & 1.560 & 13.70 & 0.736 & 0.551 & 4.015 & 978.9  & 0.470 & 0.168 \\ \hline
7 & 1.603 & 14.19 & 0.692 & 0.596 & 3.830 & 1041.7 & 0.575 & 0.167 \\ \hline
\hline
\end{tabular}
\caption{The properties of the reference stellar models. The models, obtained
using the equation of state SLy4, correspond to a
sequence of constant total central energy density $\varepsilon_{\rm c} = 9.4769 \times 10^{14}$ g/cm$^{3}$
stars with increasing rotational frequency.
From left to right, the columns represent
the gravitational mass $M$, the equatorial radius $R_{\rm eq}$,
the polar-to-equatorial radius ratio $\mathfrak{r}$,
the dimensionless angular momentum $j \equiv c J / (G M^2)$, the dimensionless
quadrupole moment $q \equiv - c^4 Q / (G^2 j^2 M^3)$, the rotational frequency $f$ and
the compactness and spin parameter duo $\kappa$, $\sigma$.
The surfaces of models 1, 3, 5 and 7 are shown in the right panel of Fig.~\ref{fig:mass_radius}.
}
\label{tab:ref_stars}
\end{table}

\begin{figure}[t!]
\includegraphics[width=\columnwidth]{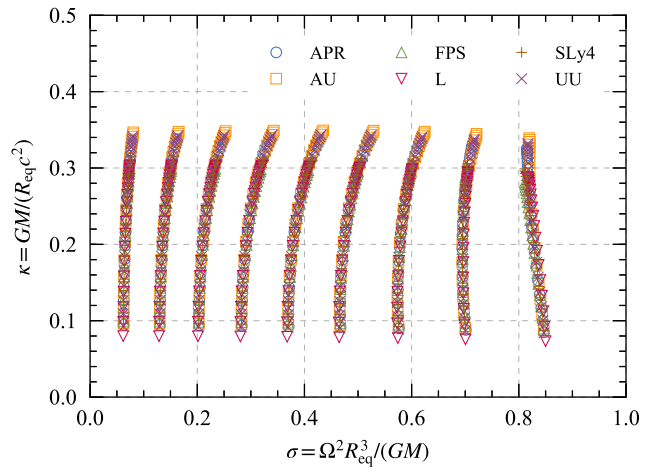}
\caption{Neutron star models parametrized by $\sigma$ and $\kappa$
used in our analytical fits to model the stellar surface. We used
an equation of state catalog that covers a wide range of stiffness.
Models with fixed $\mathfrak{r}$ lie on approximately equation-of-independent-curves
in this plane.
}
\label{fig:parameter_space}
\end{figure}

In Fig.~\ref{fig:parameter_space} we show our complete set of neutron star
solutions. It is convenient to show it not in the usual mass-(equatorial) radius
plane, but instead in a plane spanned by the dimensionless parameters,
\begin{subequations}
\begin{align}
\kappa &\equiv \frac{GM}{R_{\rm eq} c^2}\,,
\,\,\quad\textrm{(compactness)}
\\
\sigma &\equiv \frac{\Omega^2 R^{3}_{\rm eq}}{GM}\,,
\quad\textrm{(dimensionless spin parameter)}
\end{align}
\end{subequations}
where $\Omega = 2 \pi f$ is the angular frequency and we
momentarily restored factors of $c$ and $G$.
In this parametrization, the neutron star solutions
fall on approximately equation-of-state-independent curves, whose location
depends on $r_{\rm pol} / r_{\rm eq}$.
If we loosely base our definition of a rapidly rotating neutron star
as one with $\sigma \gtrsim 0.2$, which is approximately the value of $\sigma$
for a canonical neutron star $1.4\, {\rm M}_{\odot}$
(described by equation of state SLy4) spinning at $f \approx 716$~Hz
(the frequency of the fastest known pulsar to date~\cite{Hessels:2006ze})
we see that the largest fraction of our catalog consist of rapidly rotating stars.
We include very rapidly rotating stars in our work precisely because we want to
study how sensitive the fitting functions that NICER uses are to their targets
being slowly rotating. Currently, all NICER targets are indeed (relatively)
slowly rotating (with rotational frequencies smaller than
$300$~Hz~\cite{Bogdanov:2019ixe}), but in the future, it may be the case that
more rapidly rotating targets are found.

What would a typical value of $\sigma$ for a NICER target be? To answer this
question, we used the Markov-Chain Monte Carlo samples obtained by the
Illinois-Maryland analysis of the millisecond pulsar PSR~J0030+0451~\cite{miller_m_c_2019_3473466,Miller:2019cac},
which has a known rotation frequency of $f=205.53$~Hz~\cite{Lommen:2000yt,Arzoumanian:2017puf}.
We found that the best fit value to be $\sigma = 0.02$, indicating that
PSR~J0030+0451 is \emph{very} slowly rotating in the sense described above. In
this regime, neutron stars can be very well-described with the Hartle-Thorne
formalism~\cite{Hartle:1967he,Hartle:1968si,Berti:2004ny}.
In general, this formalism cannot be used to describe rapidly rotating stars,
which thus forces us to rely on numerical codes such as RNS.

\subsection{Locating the surface}
\label{sec:how_to_fit}
Having obtained a numerical neutron star solution with RNS, how do we locate
its surface?  To do this, we take advantage of the first integral of the
equation of hydrostationary equilibrium. The equation of
hydrostationary equilibrium for a uniformly rotating star with constant
angular velocity $\Omega$ is~\cite{Friedman:2013xza}
\begin{equation}
\frac{\nabla_a\, p}{(\varepsilon+p)}=\nabla_a\ln u^t\,,
\label{eq:hydroeq}
\end{equation}
where $u^a=u^t \, (t^a+\Omega\,\phi^a)$ is the 4-velocity of a fluid element
expressed in terms of the timelike and spacelike Killing vectors $t^a$ and
$\phi^a$ respectively, while
\begin{equation}
u^t= \frac{\exp\left[-(\rho+\gamma)/2 \right]}{\sqrt{1-(\Omega-\omega)^2r^2\sin^2\theta\exp(-2\rho)}}\,,
\label{eq:u_t}
\end{equation}
which follows from the normalization condition $u^{a} u_{a} = -1$ and the
line element in Eq.~\eqref{eq:line_element}.

For a barotropic equation of state, that is, one where the energy density
$\varepsilon$ and pressure $p$ are related as $\varepsilon=\varepsilon(p)$, if
one defines the enthalpy per unit mass as
\begin{equation}
h(p) \equiv \int_0^p \frac{\dd p'}{(\varepsilon+p)}\,,
\end{equation}
a first integral of Eq.~\eqref{eq:hydroeq} is
\begin{equation}
h - \ln u^t = \textrm{const.} =\left( \frac{\rho+\gamma}{2}\right)_{\rm pol}\,,
\label{eq:enthalpyInt}
\end{equation}
where the right-hand side of the equation is evaluated at the pole of the star
$r_{\rm pol} = r(\mu = 1)$.
One can verify that at the surface of the star on the pole, the enthalpy goes
to zero and it is zero along the entire surface of the star, while it is
\emph{positive in the interior} and \emph{negative in the exterior} of the
star.

\begin{figure}[t]
\includegraphics[width=\columnwidth]{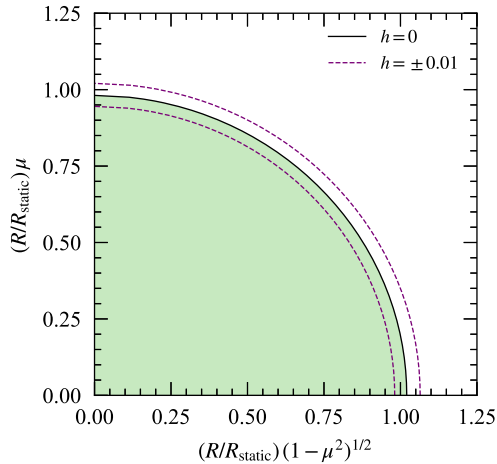}
\caption{Contours of constant enthalpy per unit mass ($h$) for Model 1 of
Table~\ref{tab:ref_stars}. We depict contours of constant $h = \pm 0.01$ near the surface.
The contour inside the stellar surface has $h>0$ and the contour outside has
$h<0$.
The surface $R_s(\mu)$ of the star at $h=0$ is indicated by the solid line.
The axes correspond to $(R / R_{\rm static}) \, (1-\mu^2)^{1/2}$ and $(R / R_{\rm static}) \, \mu$,
where $R_{\rm static} = 11.76$~km, as in the right panel of Fig.~\ref{fig:mass_radius}.
}
\label{fig:surface_locate}
\end{figure}

The RNS code provides the value of the polar redshift,
\begin{equation}
z_{\rm pol} = \exp\left[-(\rho_{\rm pol} + \gamma_{\rm pol})/2 \right]-1\,,
\end{equation}
and the surface can then be found from the condition that $h=0$ at $r = r_{\rm s}$, where
the constant in~\eqref{eq:enthalpyInt} is $-\ln(1 + z_{\rm pol})$.
Next, using Eq.~\eqref{eq:u_t}, we solve the equation
\begin{equation}
u^{t}(r, \mu) - z_{\rm pol} - 1 = 0 \,,
\label{eq:surf_condition}
\end{equation}
searching, in a sequence of values of $\mu \in [0,1]$, for the values of $r$
such that~\eqref{eq:surf_condition} is satisfied.
This gives $r_{\rm s}$ and then we can find
the circumferential radius using Eq.~\eqref{eq:circ_radius}.
A {\sc{Mathematica}} notebook implementing these steps can be found in~\cite{GPRNSRepo}.
As an example we show in Fig.~\ref{fig:surface_locate} the contours of
constant enthalpy per unit mass $h$ for Model 1 in Table~\ref{tab:ref_stars}. The
surface is indicated with a solid line, which corresponds to $h=0$, while the
two dashed lines correspond to $h = \pm 0.01$.

\subsection{Analytical fits}
\label{sec:fit_formulas}

Having obtained the data for the surface $R(\mu)$ of each star in our ensemble,
we can now obtain analytical fits that describe the surfaces of {\it all}
stars.
The procedure to generate such a formula is simple and was first explored
in~\cite{Morsink:2007tv}: we first fit a proposed formula $R(\mu; \{a_{n}\})$ (that
depends on one or more free constants $a_{n}$) for each star, parametrized by
its compactness $\kappa$ and spin parameters $\sigma$ (see Fig.~\ref{fig:parameter_space}).
The outcome of this procedure is a table $\{a_{n}, \, \kappa, \, \sigma \}$.
This data can then be fitted to some analytical representation
$a_{n} (\kappa,\, \sigma)$.
These steps result in
a formula $R(\mu; \{\kappa,\,\sigma\})$ for the surface.

We stress that this process introduces a \textit{smearing} of the particular
way in which deformations away from sphericity take place for neutron stars
described by different equations of state as the rotation frequency increases.
For practical applications, such as pulse profile modeling, but see
also for the cooling tail method~\cite{Suleimanov:2016llr,Suleimanov:2020ijb}, an ideal formula
$R(\mu)$ would capture accurately the neutron star surfaces at a wide range of
spin frequencies, compactness and a wide set of equations of state (i.e. it has
to be quasi-equation-of-state independent~\cite{Morsink:2007tv,AlGendy:2014eua,Yagi:2016bkt}).

In the remainder of this subsection, we review two formulas used in the literature
(Secs.~\ref{sec:morsink_etal_fits} and~\ref{sec:algendy_etal_fits}) that share
these properties and also introduce a new formula (Sec.~\ref{sec:eccentricity_fits}).

\subsubsection{The \ms formula}
\label{sec:morsink_etal_fits}

In~\cite{Morsink:2007tv}, \ms introduced a formula
based on the assumption that the surface is related
with the equatorial radius $R_{\rm eq}$ as
\begin{equation}
R_{\rm M}(\mu) = R_{\rm eq}
\left[1 + \sum_{n = 0}^{2} a_{2n}(\sigma,\, \kappa) P_{2n}(\mu)\right]\,,
\label{eq:fit_morsink_etal}
\end{equation}
where $P_{\ell}(\cdot)$ are Legendre polynomials, $\mu = \cos\theta$ and
$a_{2n}$ are coefficients that depend on both $\sigma$ and $\kappa$ as
\begin{equation}
a_{2n} \equiv c_{(1,0)} \sigma + c_{(1,1)} \sigma \kappa + c_{(2,0)} \sigma^2\,,
\label{eq:coef_morsink}
\end{equation}
where $c_{(i,j)}$ is the coefficient multiplying the product $\sigma^{i} \kappa^{j}$.
This notation will be used throughout this work.

The argument $\mu$ for the Legendre polynomials is chosen to enforce the $\mathbb{Z}_2$-symmetry
of the star's surface across the equator and the even-order Legendre polynomials
are used to force that $R_{\rm M}(\mu) = R_{\rm M}(-\mu)$ across
the spin axis.
Up to $n=1$, this formula corresponds to the first
order rotation-induced deformations in Hartle's perturbative
expansion~\cite{Hartle:1967he}, while the $n=2$ term
captures higher-order spin deformations of the star\footnote{In principle,
one could work within the Hartle-Thorne formalism beyond second-order in spin
to study the surface semianalytically. See~\cite{Benhar:2005gi} for the extension to
third-order in spin and~\cite{Yagi:2014bxa} for the fourth-order in spin calculation.
For pulse profile calculations in Hartle-Thorne spacetimes see~\cite{Psaltis:2013zja,Oliva-Mercado:2020nnk}.}.
In the nonrotating limit ($\sigma = 0$), we have $a_{2} = a_{4} = 0$ and
therefore $R_{\rm M} = R_{\rm eq}$ for all $\mu$.
One caveat of Eq.~\eqref{eq:fit_morsink_etal} is that it does not
satisfy the consistency condition $R_{\rm M}(0) = R_{\rm eq}$.
However, the mismatch between $R_{\rm M}(1)$ and $R_{\rm eq}$
is less than 1\%~\cite{Morsink:2007tv}.

The coefficients $c_{(i,j)}$ in Eq.~\eqref{eq:coef_morsink} are summarized
in Table~\ref{tab:coefs_ma}.
For self-consistency in our analysis, we recalculated the values of these
coefficients using our neutron star ensemble, which differs from
that used in~\cite{Morsink:2007tv} in size, rotation frequencies sampled
and equations of state used.
The values quoted between parenthesis in Table~\ref{tab:coefs_ma} correspond
to the values found in~\cite{Morsink:2007tv}. We see that in general our values
are in good agreement.

\begin{table*}[t]
\begin{tabular}{ l | c  c  c}
	\hline
	\hline
    Surface model & & Coefficient & \\
	\hline
    & $c_{(1,0)}$ & $c_{(1,1)}$ & $c_{(2,0)}$ \\
    \ms \cite{Morsink:2007tv} & & & \\
	$a_{0}$ & $(-0.18) \, -0.193$ & $(+0.23) \, +0.092$ & $(-0.05) \, -0.015$ \\
	$a_{2}$ & $(-0.39) \, -0.391$ & $(+0.29) \, +0.088$ & $(+0.13) \, +0.149$ \\
	$a_{4}$ & $(+0.04) \, +0.031$ & $(-0.15) \, -0.064$ & $(+0.07) \, +0.086$ \\
	\hline
	\ag \cite{AlGendy:2014eua} & & & \\
	$a_{2}$ & $(-0.788) \, -0.533$ & $(+1.030) \, +0.203$  & - \\
	\hline
	\hline
\end{tabular}
\caption{Values of the coefficients $c_{(i,j)}$ for \ms and \ag fits.
The values quoted in
parenthesis correspond to the original values found in~\cite{Morsink:2007tv,AlGendy:2014eua},
obtained using a neutron star ensemble different from ours.
We attribute the larger differences in the \ag fitting coefficients
between our values and the ones quoted in~\cite{AlGendy:2014eua}
to the fact those were obtained using a ensemble of mostly slowly rotating stars with
spin parameter $\sigma \leq 0.1$, whereas we do include rapidly rotating models with $\sigma$
as large as $0.8$ (see Fig.~\ref{fig:mass_radius}).
}
\label{tab:coefs_ma}
\end{table*}

\begin{table*}[t]
\begin{tabular}{ l | c  c  c  c  c  c  c}
	\hline
	\hline
    Surface model &   & & & Coefficient & & &\\
	\hline
    & $c_{(0,0)}^{(y)}$ & $c_{(1/2,0)}^{(y)}$ & $c_{(1,0)}^{(y)}$ & $c_{(0,1)}^{(y)}$ & $c_{(1,1)}^{(y)}$ & $c_{(2,0)}^{(y)}$ & $c_{(0,2)}^{(y)}$\\
	{{Slow-elliptical fit}} &  \\
    $y=e$     & - & $+1.089$ & $+0.168$ & - & $-0.685$ & $-0.802$ & - \\
    $y=a_{2}$ & $-1.013$ & - &     $-0.312$ & - & $+0.930$ & $-1.596$ & - \\
    $y=a_{4}$ & $+0.016$ & - &     $+0.301$ & - & $-1.261$ & $+2.728$ & - \\
    % ---------------------
    \hline
	{{Fast-elliptical fit}} &  \\
    $y=e$     & $+0.251$ & - & $+0.935$ & $+0.709$ & $+0.030$ & $-0.472$ & $-2.427$ \\
    $y=a_{2}$ & $-1.265$ & - & $+0.220$ & $+2.651$ & $+1.010$ & $-1.815$ & $-7.657$ \\
    $y=a_{4}$ & $+0.556$ & - & $-1.465$ & $-4.260$ & $-2.327$ & $+4.921$ & $+12.98$ \\
	\hline
	\hline
\end{tabular}
\caption{Values of the coefficients $c_{(i,j)}$ for the slow and fast variations
of the elliptical fit.
The former only uses stars for which the spin parameter is $\sigma \leq 0.25$, while the latter only
those for which $\sigma \geq 0.2$.
}
\label{tab:coefs_el}
\end{table*}

\subsubsection{The \ag formula}
\label{sec:algendy_etal_fits}

An alternative to Eq.~\eqref{eq:fit_morsink_etal} that satisfies the
constraint $R(0) = R_{\rm eq}$ was proposed by \ag~\cite{AlGendy:2014eua}
\emph{and is currently in use in the pulse profile modeling by NICER}~\cite{Bogdanov:2019qjb}.
Their formula is
\begin{align}
R_{\rm A}(\mu) &= R_{\rm eq}
\left[
1 - \left(1-\mathfrak{r}\right)\, \mu^{2}
\right] \equiv R_{\rm eq}
\left[
1 + a_{2}(\kappa, \sigma)\, \mu^{2}
\right]\,,
\nonumber
\\
\label{eq:fit_algendy_etal}
\end{align}
where the coefficient $a_{2}$ is given by
\begin{equation}
a_{2} = c_{(1,0)} \sigma + c_{(1,1)} \sigma \kappa \,,
\end{equation}
and represents the multiplicative factor $(1-\mathfrak{r})$,
which contains both the equatorial $R_{\rm eq}$ and polar $R_{\rm pol}$ radii
of the star [cf.~Eq.~\eqref{eq:ratio_poleq}].
Due to the same symmetry requirements as in the \ms fit, even powers
of $\mu = \cos\theta$ are used.
In the nonrotating limit $(\sigma = 0)$, $a_{2} = 0$ and therefore $R_{\rm A} = R_{\rm eq}$
for all $\mu$.

The values of $c_{(1,0)}$ and $c_{(1,1)}$ are quoted in Table~\ref{tab:coefs_ma}.
As we did previously for the \ms formula, we recalculated the fitting
coefficients using our own neutron star ensemble.
We find larger differences between our values and those quoted
in~\cite{AlGendy:2014eua}.
We credit these differences due to the fact that Ref.~\cite{AlGendy:2014eua}
only considered slowly rotating stars ($\sigma \leq 0.1$) whereas our catalog
consists of mostly rapidly rotating stars ($\sigma \geq 0.25$), as we have
discussed before.

\subsubsection{The elliptical formula}
\label{sec:eccentricity_fits}

In addition to the models previously described, we also introduce
a \emph{new} expression. Our choice is inspired by the elliptical isodensity
approximation~\cite{Lai:1993ve} and is given by:
\begin{equation}
R_{\rm E}(\mu) = R_{\rm eq}
\sqrt{\frac{1-e^2}{1-e^2 \, g(\mu)}}\,,
\label{eq:fit_elliptic}
\end{equation}
where
\begin{align}
g(\mu) &= 1 + a_{2}(\kappa,\sigma) \, \mu^2
        + a_4(\kappa,\sigma) \, \mu^4 \nonumber \\
        &\quad - [1 + a_2(\kappa,\sigma) + a_{4}(\kappa,\sigma)] \, \mu^6 \,.
\label{eq:gfun}
\end{align}
and the term multiplying $\mu^6$ was chosen such that,
\begin{equation}
\frac{R_{\rm E}(1)}{R_{\rm E}(0)} = \mathfrak{r} = \sqrt{1 - e^2}\,,
\label{eq:consistency}
\end{equation}
thereby enforcing the interpretation of $e$ as the \textit{star's eccentricity}~\cite{Stein:2013ofa}.
As in the previous fitting formulas, even powers of $\mu$ are used to
enforce $R_{\rm E} (\mu) = R_{\rm E}(-\mu)$.
At a qualitative level our formula differs from Eqs.~\eqref{eq:fit_morsink_etal}
and~\eqref{eq:fit_algendy_etal} in that we are \textit{including relativistic
and spin corrections to an otherwise ellipsoidal star}, whereas the other two fits
are \textit{including relativistic and spin corrections to an otherwise spherical
star}. Using an ellipsoidal star as the unperturbed configuration is motivated by
the fact that in Newtonian gravity rotating stars are not spheres, but rather they
are ellipsoids of revolution.

We obtained \emph{two} fits using our elliptic formula. The first, which
we name the \emph{``slow elliptical''} fit, uses only stars with
$\sigma \leq 0.25$. The second, which we name the \emph{``fast elliptical''}
fit, uses only stars with $\sigma \geq 0.2$.
The reasons are twofold.
First, on the observational side, the fastest known millisecond pulsar has a
frequency of 716~Hz~\cite{Hessels:2006ze}, which is approximately 2.5 times
the rotation frequency of the fastest spinning NICER's target~\cite{Bogdanov:2019ixe},
PSR~J1231--1411 which has a rotation frequency of $271.7$~Hz~\cite{Ransom:2010if}.
Second, on the practical side, the majority of the stars in our catalog have
$\sigma > 0.25$, which corresponds approximately to minimum rotation frequencies
in the 700-800~Hz range.
Therefore, any fit obtained using the full catalog will be \emph{skewed}
toward the values of coefficients $c_{(i,j)}$ corresponding to rapidly
rotating stars.
These two observations suggest separating our fits in the slow and fast fits,
including a ``buffer $\sigma$-region'' where they overlap.

The coefficients $e$, $a_{2}$ and $a_{4}$ are determined by
\begin{align}
y &=
c_{(0,0)}^{(y)}
+ c_{(1/2, 0)}^{(y)} \sigma^{1/2}
+ c_{(1,0)}^{(y)} \sigma
+ c_{(0,1)}^{(y)} \kappa
\nonumber \\
&\quad + c_{(1,1)}^{(y)} \sigma \kappa
+ c_{(2,0)}^{(y)} \sigma^2
+ c_{(0,2)}^{(y)} \kappa^2
\,,
\end{align}
with $y$ any of $e$ or $a_{2n}$.
In the slow-elliptical fit we set
\begin{subequations}
\begin{align}
c_{(0,0)}^{(e)} = c_{(1,0)}^{(e)} = c_{(2,0)}^{(e)} = 0\,,
\\
c_{(1/2,0)}^{(a_{2n})} = c_{(1,0)}^{(a_{2n})} = c_{(2,0)}^{(a_{2n})} = 0\,,
\end{align}
\end{subequations}
since to impose the nonrotating limit we must set all $\sigma$-free coefficients to zero.
The peculiar fractional-order coefficient $c_{(1/2,0)}$ is introduced to capture better the
behavior of the eccentricity $e$ in the $\sigma \ll 1$ limit.
As for the fast-elliptical fit, we do not need to impose these restrictions on the $\sigma$-free
coefficients, but we do set
\begin{equation}
c_{(1/2,0)}^{(y)} = 0, \quad \textrm{for} \quad \, y = \{e,\,a_{2n} \}\,,
\end{equation}
since its introduction was motivated by $e$ in the small-$\sigma$ limit.
The coefficients $c_{(i,j)}$ for both flavors of the elliptic fit are
summarized in Table~\ref{tab:coefs_el}.

\subsection{Comparison between the different formulas}
\label{sec:fit_comparison}

% \subsubsection{``Exact'' surfaces}
% \label{sec:exact_surf}

In the previous section, we introduced three formulas that describe the surface of
neutron stars for a wide range of spin and compactness parameters.
How do they compare when confronted against the properties of
\textit{individual} neutron star models computed as accurately as possible?
Neutron stars are generally believed
to be described by a \textit{single equation of state}.
Therefore, using fits which integrate out the surface variability of neutron
stars due to different equations of state could introduce a source of systematic
error in any neutron star parameter estimation where the fits are used.

As a first step to analyze this source of systematic error, in this section we
compare the three formulas (using our own fitting coefficients)
against neutron star models computed numerically
with the equation of state SLy4~\cite{Douchin:2001sv}.
We use our own fitting coefficients for all three formulas to avoid a
systematic error introduced by comparing different fits obtained from
different neutron-star catalogs. Recall that the catalogs used here and
in Refs.~\cite{Morsink:2007tv,AlGendy:2014eua} are all different.
We chose the equation of state SLy4 because it yields neutron stars with masses
greater than $1.9 \, {\rm M}_{\odot}$ as required by the observations of the
massive pulsars
J1614$-$2230~\cite{Demorest:2010bx,Fonseca:2016tux,Arzoumanian:2018saf},
J0348+0432~\cite{Antoniadis:2013pzd}
and
J0740+6620~\cite{Cromartie:2019kug},
and yet it is relatively soft as required by tidal deformability estimates from
the GW170817 event~\cite{TheLIGOScientific:2017qsa,Abbott:2018exr,Abbott:2018wiz}.

Let us first describe the neutron star models we will use as benchmarks in this
section and in the remainder of this work. We use a sequence of stars parametrized by
their central energy density $\varepsilon_{\rm c}$ ($= 9.4769 \times 10^{14}$ g/cm$^3$),
which for the SLy4 equation of state results results in a ``canonical'' neutron
star with a mass of approximately $1.4\,{\rm M}_{\odot}$ in the nonrotating
limit.
The properties of these ``benchmark stars'' are summarized in Table~\ref{tab:ref_stars}
and they are indicated by markers in the mass-(equatorial) radius plane in Fig.~\ref{fig:mass_radius}.
We will use the term ``benchmark'' to any property or observable calculated using one of these stars.
For instance, we will refer to their surfaces as ``benchmark surfaces'' and to
the pulse profile emitted from their surface as ``benchmark pulse profiles.''

To describe the shape of these stars as accurately as possible we fit
\emph{separately} both $R = R(\mu$) and
\begin{equation}
\frac{\dd \log R(\mu)}{\dd \theta} = - (1-\mu^2)^{1/2} \frac{1}{R(\mu)} \frac{\dd R(\mu)}{\dd\mu}\,,
\label{eq:log_deriv_definition}
\end{equation}
the latter
being a measurement of the deviation from sphericity of the star's surface
and subject to the constraints
\begin{equation}
[\dd \log R(\mu) / \dd \theta]_{\mu=0}
= [\dd \log R(\mu) / \dd \theta]_{\mu=1}
= 0\,.
\label{eq:oblate_constraints}
\end{equation}

\begin{table*}[!htpb]
\begin{tabular}{ c | c c c c c | c  c | c c c }
	\hline
	\hline
    Model & $10^{-2} \cdot a_{2} $ & $10^{-2} \cdot a_{4}$ & $10^{-2} \cdot a_{6}$ & $10^{-2} \cdot a_{8}$ & $10^{-2} \cdot a_{10}$
          & $10^{-2} \cdot b_{1}$  & $10^{-2} \cdot b_{3}$
          & $10      \cdot c_{0}$  & $10      \cdot c_{2}$ & $10      \cdot c_{4}$ \\
    \hline
    1 & $-4.174$ & $0.3647$ & $-0.0495$ & $0.01175$ & $-0.0034$
      & $-5.213$ & $-5.623$
      & $6.244$ & $7.575$ & $0.811$
	\\ \hline
    2 & $-9.132$ & $1.832$ & $-0.531$ & $0.1603$ & $-0.0310$
      & $-10.42$ & $ -10.67$
      & $5.702$ & $ 7.656$ & $1.464$
	\\ \hline
    3 & $-15.15$ & $5.349$ & $-2.644$ & $1.173$ & $-0.276$
      & $-18.16$ & $-12.09$
      & $5.992$ & $7.373$ & $1.403$
	\\ \hline
    4 & $-22.68$ & $12.73$ & $-9.303$ & $5.279$ & $-1.422$
      & $-19.50$ & $-17.04$
      & $4.295$ & $7.769$ & $1.895$
	\\ \hline
    5 & $-32.44$ & $27.49$ & $-27.20$ & $18.32$ & $-5.423$
      & $-17.76$ & $-23.03$
      & $2.724$ & $7.710$ & $2.603$
	\\ \hline
    6 & $-45.69$ & $56.56$ & $-70.98$ & $54.16$ & $-17.17$
      & $-16.07$ & $-30.44$
      & $1.731$ & $7.760$ & $3.507$
	\\ \hline
    7 & $-64.72$ & $114.0$ & $-172.8$ & $145.0$ & $-48.43$
      & $-13.76$ & $-38.70$
      & $1.013$ & $7.644$ & $4.591$
	\\ \hline
	\hline
\end{tabular}
\caption{Values of the coefficients $a_{i}$ in the fitting formula
Eq.~\eqref{eq:r_exact_fit} and $b_{i}$, $c_{i}$ in Eq.~\eqref{eq:dlogr_exact_fit}
for our set of reference stellar models, whose properties are summarized in
Table~\ref{tab:ref_stars}.
}
\label{tab:ref_coefs_a}
\end{table*}

\begin{figure*}[htpb!]
\includegraphics[width=\columnwidth]{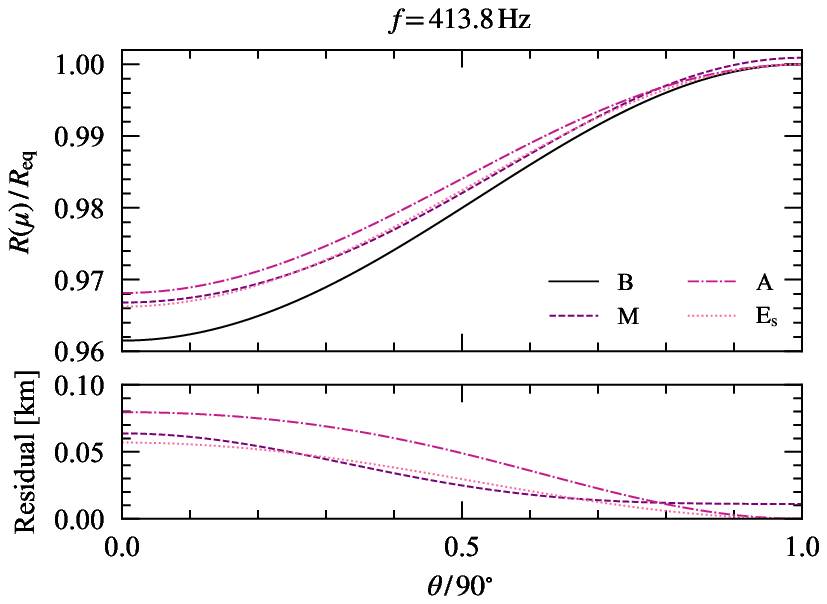}
\includegraphics[width=\columnwidth]{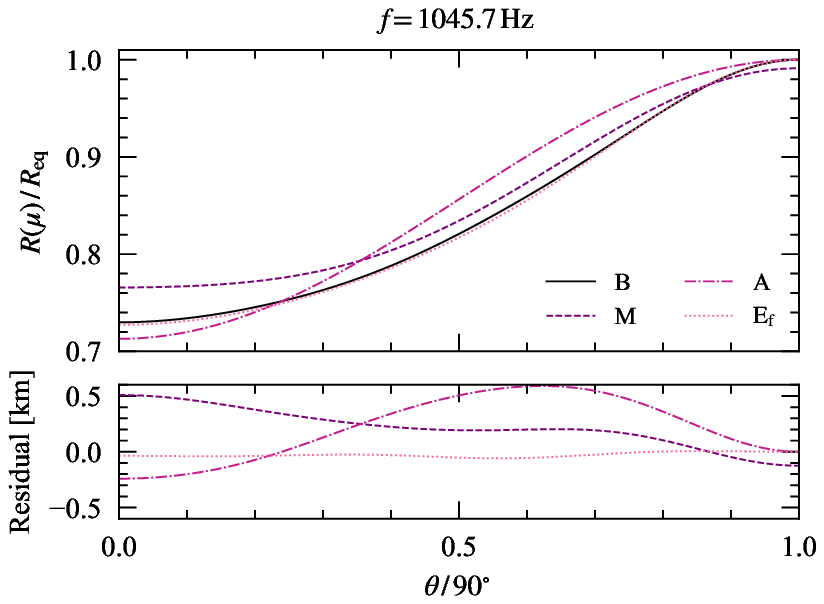} \\
\includegraphics[width=\columnwidth]{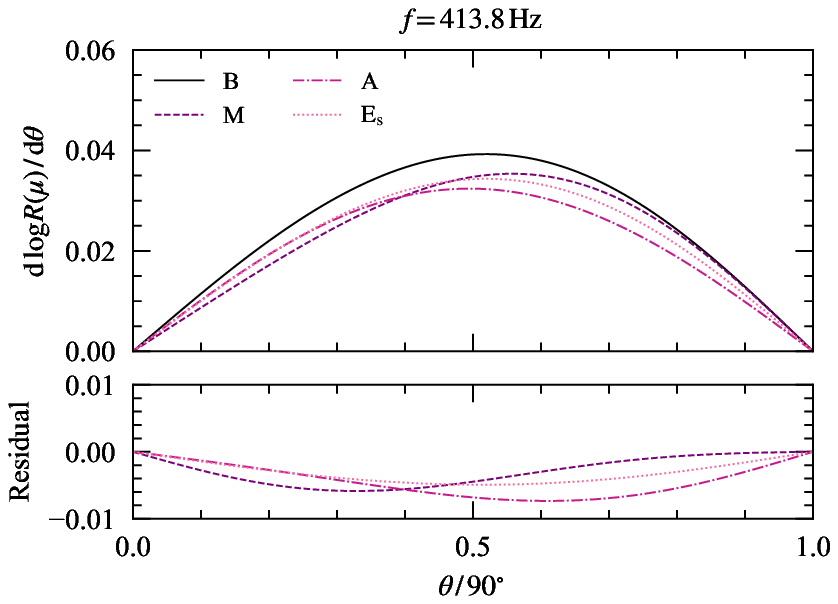}
\includegraphics[width=\columnwidth]{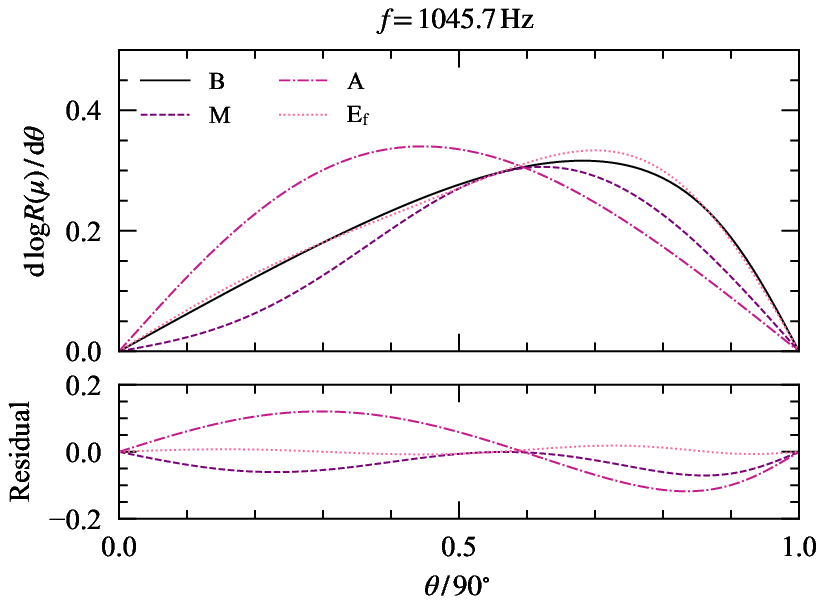}
\caption{
Surfaces (top) and surficial logarithmic derivative (bottom) of rotating neutron stars. The left-panels correspond to a star with rotation frequency of $f=413.8$~Hz, whereas
the right-panels to a star with rotation frequency of $f=1045.7$~Hz, which correspond to the benchmark stars labeled 1 and 7 in Table~\ref{tab:ref_stars}.
The top-panels show $R(\mu)$ normalized by the equatorial radius
$R_{\rm eq}$, while the bottom panels show $\dd \log R(\mu)/\dd \theta$ without normalization.
The different curves correspond to the surface as determined by the RNS code (solid line),
as predicted using the \ms fit (dashed line), the \ag fit (dash-dotted line)
and the elliptical fit (dotted lines), with the slow-elliptical fit on the left and the fast-elliptical fit on the right.
The bottom panels show the residuals between each of the fits and the benchmark surfaces.
}
\label{fig:err_surface}
\end{figure*}

For these two quantities we used the following fitting formulas
\begin{subequations}
\begin{align}
R_{\rm B}(\mu) &= R_{\rm eq}\left(1 + \sum_{k=1}^{5} a_{2k} \mu^{2k}\right)\,,
\label{eq:r_exact_fit}
\\
\frac{\dd \log R_{\rm B}(\mu)}{\dd \theta} &=
- (1-\mu^2)^{1/2}
\frac{\sum_{k=0}^{2} b_{2k+1} \mu^{2k+1}}{\sum_{k=0}^{3} c_{2k} \mu^{2k} }\,.
\label{eq:dlogr_exact_fit}
\end{align}
\label{eq:exact_fits}
\end{subequations}
\noindent Equation~\eqref{eq:r_exact_fit} is a higher-order \ag fit,
with the higher powers of $\mu$ introduced to describe the
greater deformations away from spherical symmetry that happen at
high rotation frequencies and to simultaneously retain the property that $R_{\rm B}(0) = R_{\rm eq}$.
Equation~\eqref{eq:dlogr_exact_fit} is chosen such as to represent the
logarithmic derivative of Eq.~\eqref{eq:r_exact_fit} and, by construction,
it satisfies the constraints of Eq.~\eqref{eq:oblate_constraints}.

We used these formula to fit our numerical data and the resulting fitting coefficients
$a_{i}$, $b_{i}$ and $c_{i}$ are summarized in Table~\ref{tab:ref_coefs_a}.
To obtain the fits for Eq.~\eqref{eq:dlogr_exact_fit}, we first calculated numerically
the logarithmic-derivative using a sixth-order finite difference formula.
A detailed study of the numerical derivatives and the goodness of the fits is
presented in Appendix~\ref{app:exact}.

In the top panel of Fig.~\ref{fig:err_surface} we show the residuals~$R_{\rm fit} - R_{\rm B}$,
as functions of the colatitude $\theta$, between the three fitting formulas and the benchmark stars for the slowest and fastest
rotating stars in Table~\ref{tab:ref_stars}.
We see that for the slowest rotating model (left-panel) the \ms
and the (slow) elliptic fit behave very similarly and they are both closer to the benchmark
surface in comparison to the \ag formula.
Nonetheless, the residuals are small, below $0.1$~km, indicating that all three
formulas agree well with the benchmark surface.
The situation changes when we consider the fastest rotating model (right-panel).
We see that the (fast) elliptical fit outperforms both the \ms and the \ag fits.
For the latter two formulas the largest value of the residual increases approximately fivefold,
however, staying bound to be less than 0.5~km.
In the bottom panel of Fig.~\ref{fig:err_surface} we carry out the same analysis but for the
logarithmic-derivative of the surface, reaching similar conclusions.

\section{Implications of the fitting formulas on the pulse profiles}
\label{sec:lightcurves}

In the previous section we introduced the various fitting expressions for the
surface of rotating neutron stars and studied how well they reproduce a set of
benchmark surfaces.
How does the mismatch between fit and benchmark surfaces appear
in the pulse profile generated by hot spots on the star's surface?
In this section we address this question in two fronts.
First, given that the surface depends on the colatitude $\theta$,
it is clear that the mismodeling of pulse profiles
will depend both on where on the surface the hot spot is located
($\theta_{\rm s})$ and on the line of sight of the observer ($\iota_{\rm o}$),
where both angles measured relative to the rotation axis of the star.
Therefore, it is natural to examine for which combinations
$(\theta_{\rm s},\,\iota_{\rm o})$ the mismatch is smallest/largest.
Second, we want to explore how the different formulas perform
when trying to extract the equatorial radius $R_{\rm eq}$ from
a synthetic injection.
Of course, both questions are intertwined as, for instance,
a combination $(\theta_{\rm s},\, \iota_{\rm o})$ for which the
flux mismatch is large will, likely, result in a large systematic
error in the inference of $R_{\rm eq}$.
For the reasons explained in Sec.~\ref{sec:fit_comparison}, we continue to use
the surface formulas with our own set of fitting coefficients.

To answer these questions we need to construct (as accurately
as possible) reference pulse profiles to compare against.
Ideally, these ``benchmark pulse profiles'' should be calculated doing
ray-tracing on a numerically constructed rotating neutron star spacetime.
For simplicity, we restrict ourselves to the O+S model, with the star's
oblateness modeled by the high-order fitting expressions introduced in
Sec.~\ref{sec:fit_comparison}.

As already mentioned, the O+S model is currently used by NICER and its validity has extensively been
examined by comparison against ray-tracing in numerically obtained spacetimes
of rotating neutron stars.
These studies have shown that the O+S model can accurately describe the x-ray
emission of the neutron star surfaces for a typical NICER target.
Our own implementation of the O+S model follows closely the presentation in
Refs.~\cite{Morsink:2007tv,Salmi:2018gsn,Bogdanov:2019qjb}.
The code was validated against the Alberta code described
in~\cite{Bogdanov:2019qjb} which, in turn, has been validated against several
other codes used in the NICER collaboration.

In all calculations in this work, we assume for simplicity a pointlike hot spot
with angular radius $\Delta \theta_{\rm s} = 0.01^{\circ}$.
We further assume that this hot spot radiates isotropically according
to a blackbody spectrum with $k_{\rm B} T'_0 = 0.35$~keV
(measured by an observer comoving with the hot spot).
We place the observer at a distance $d = 200$~pc from the source and we assume
that this observer collects photons arriving with $E = 1$~keV.
We also fix the initial phase of the observed flux (i.e. its zero value)
to when the hot spot is closest to the observer.
This is a representative value within the soft x-ray band in which NICER
operates. These quantities are summarized in Table~\ref{tab:lightcurve_params}.
\begin{table}[t]
\begin{tabular}{l  c}
\hline
\hline
Parameter & Value \\
\hline
Hot spot angular radius ($\Delta \theta_{\rm s}$) & $0.01$~deg \\
Hot spot temperature at comoving frame ($k_{\rm B} T'_0$) & $0.35$~keV \\
Observed photon energy ($E$) & 1 keV \\
Distance ($d$) & 200 pc \\
\hline
\hline
\end{tabular}
\caption{Pulse profile parameters. The table summarizes the parameters that enter the
pulse profile calculation which we keep fixed throughout this work. We assume the
existence of single, pointlike hot spot on the star's surface to isolate the
effects of the different neutron star surface models on the resulting x-ray flux.}
\label{tab:lightcurve_params}
\end{table}

These simplifications allow us to \textit{isolate} the influence
of the different surface models on the pulse profiles.
However, our results must be considered \textit{conservative} since other
effects, such as the influence of frame dragging and higher-spacetime multipole
moments on the photon motion, are not taken into account in the O+S approximation.
Our analysis, while indicative of what can happen in a more complete analysis,
cannot substitute a full parameter estimation in the framework of Bayesian
inference (see, e.g.,~\cite{Lo:2013ava,Lo:2018hes,Miller:2016kae,Miller:2019cac,Riley:2019yda}),
a task which we leave for future study.

\begin{figure*}[htpb!]
\includegraphics[width=\columnwidth]{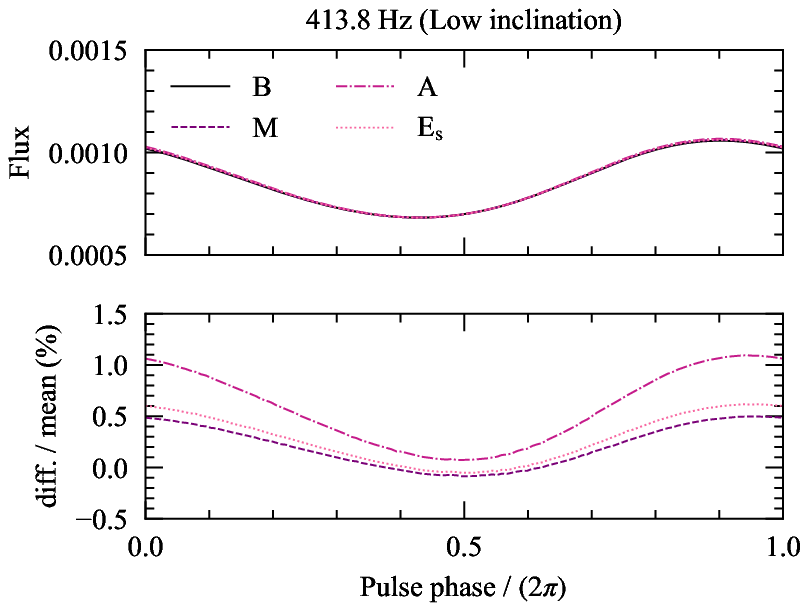}
\includegraphics[width=\columnwidth]{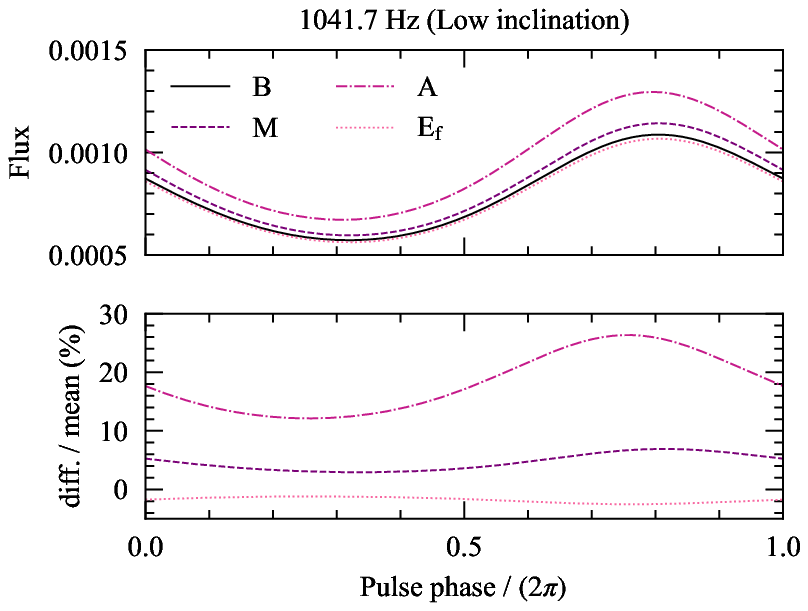} \\
\includegraphics[width=\columnwidth]{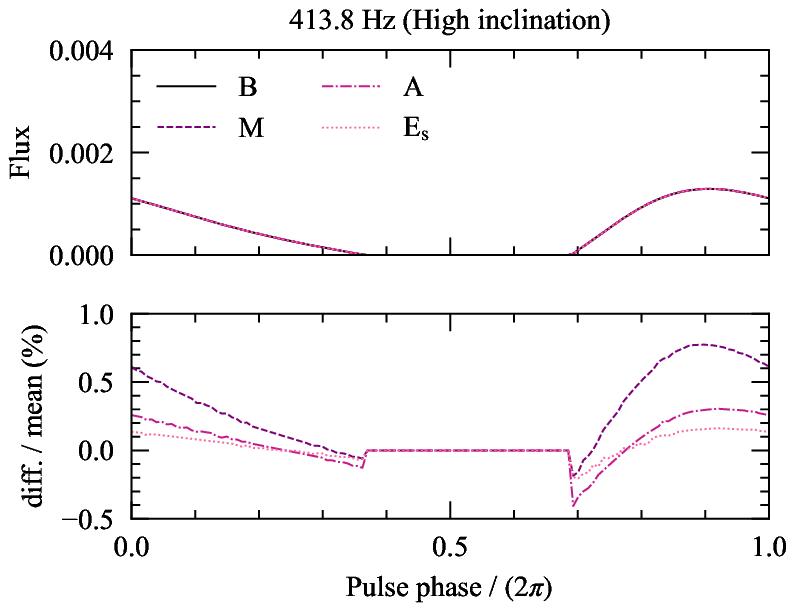}
\includegraphics[width=\columnwidth]{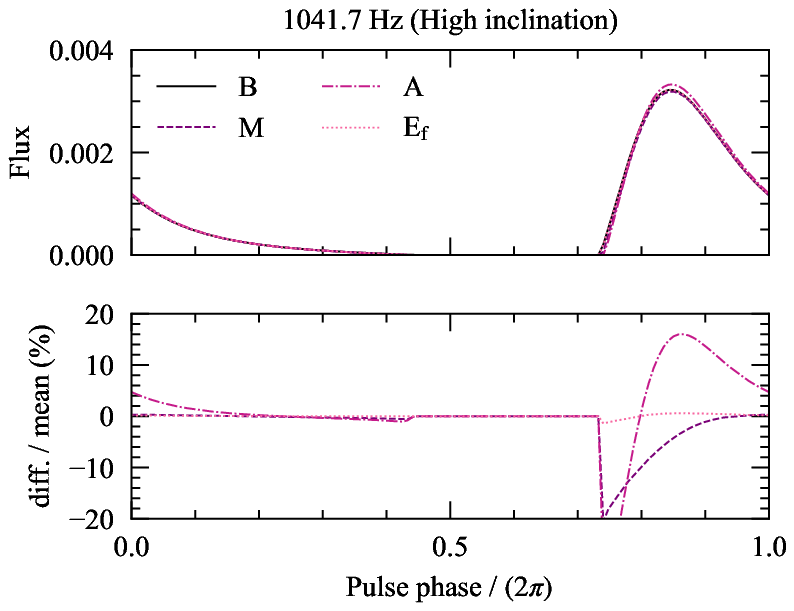}
\caption{Illustrative pulse profiles. We compare the pulse profiles, in units
of photons~cm$^{-2}$~s$^{-1}$~keV$^{-1}$, obtained using
the three different fitting formulas against those obtained by the slowest and
fastest rotating benchmark stars (models 1 and 7 in Table~\ref{tab:ref_stars}).
Top row: the low-inclination hot spot/observer orientation with the pulse profiles
in the top panel and the difference between each surface model and the benchmark pulse profiles
divided by the mean of the latter
in the bottom panel.
Bottom row: similar to the top row but for the high-inclination
hot spot-observer orientation.
}
\label{fig:example_fluxes}
\end{figure*}

In Fig.~\ref{fig:example_fluxes} we show some examples of the difference in the
pulse profile (in units of photons~cm$^{-2}$~s$^{-1}$~keV$^{-1}$) when we
fix all parameters used to produce it and only vary the fitting formula used to
model the star's surface.
We quantify this difference by subtracting the benchmark pulse profile (i.e. the one
obtained using the ``exact'' surface formula) from the pulse profile obtained using each
fitting formula and then dividing by the mean value of the former.
We consider the slowest and fastest stars in our benchmark catalog and
two hot spot-observer orientations.
The first, labeled ``low inclination''
has $(\theta_{\rm s},\, \iota_{\rm o}) = (45^{\circ},\, 20^{\circ})$,
while the second, labeled ``high inclination''
has $(\theta_{\rm s},\, \iota_{\rm o}) = (80^{\circ},\, 85^{\circ})$.
These two configurations are summarized in Table~\ref{tab:cases}.
The figure shows that for the slowest rotating model, all fitting formulas
agree with the benchmark pulse profiles with differences of at most $\sim 1\%$.
For the fastest rotating model, a larger differences appear and can be as large as $\sim 30\%$.
Except at these phase values, all formulas agree with the benchmark pulse
profile in the high-inclination case.  However, for the low-inclination case,
we see that the new fast elliptical fit does agree remarkably well with the
benchmark pulse profile.

\begin{table}[t!]
\begin{tabular}{l  c  c  c  c}
\hline
\hline
Case & $\theta_{\rm s}$ & $\iota_{\rm o}$ & $\delta{R_{\rm eq}}$ \\
     & (deg) & (deg)  &  (km) \\
\hline
Low inclination   & 45 & 20 & 0.20  \\
High inclination  & 80 & 85 & 0.05 \\
\hline
\hline
\end{tabular}
\caption{Summary of the hotspot/observer arrangements
used in the estimation of $R_{\rm eq}$ and the values
of the statistical error $\delta {R_{\rm eq}}$ used
when calculating the likelihood. The two arrangements
are located Fig.~\ref{fig:error_angle_dependence} by
the markers $\bigcirc$ (low inclination) and $\Box$ (high inclination).
The statistical errors are based on~\cite{Lo:2013ava,Lo:2018hes}.
}
\label{tab:cases}
\end{table}

\subsection{Dependence on the hot spot and observer's orientation}
\label{sec:dependence_orientation}

Let us examine the error introduced by the fitting formulas, relative
to the benchmark pulse profiles, when we vary
the hot spot ($\theta_{\rm s}$) and observer location ($\iota_{\rm o}$).
The reason for this study is the following: there is no reason for the error to be the same
for all pairs $(\theta_{\rm s}, \iota_{\rm o})$.
Indeed, as shown in Fig.~\ref{fig:err_surface}, a
surface fit can match exactly the benchmark models \emph{locally}, although
not well \emph{globally}.
If the hotspot is located at one of these special values of colatitude, the
resulting pulse profile will be the same.
The location of these ``coincident colatitudes'' depends on the frequency $f$
of the star. For instance, returning to Fig.~\ref{fig:err_surface}, we see that for the \ag
fit this happens at $\theta / 90^{\circ} = 1$ when $f = 413.8$~Hz, but at
$\theta / 90^{\circ} \approx 0.32$ and $1$ when $f=1045.7$~Hz.
An extreme example where this situation happens for all rotation frequencies
is when both $\theta_{\rm s}$ and $\iota_{\rm o}$ are on the equator
($90^{\circ}$).
In this case, as long as $R_{\rm fit} = R_{\rm eq}$ the pulse profiles will be
identical.
This happens for the \ag and elliptical fits, and to a good
approximation for the \ms formula.

We quantify the mismatch between pulse profiles predicted by the different
surface formulas over the course of a single revolution of the star using two
measures.
First, we define the mean residual
\begin{equation}
\mathscr{R} \equiv \frac{1}{N_{\rm bins}} \, \sum_{i=1}^{N_{\rm bins}} |F^{\rm B}_i - F^{\rm fit}_i|\,,
\label{eq:r_def}
\end{equation}
where $F^{\rm B}_i$ ($F^{\rm fit}_i)$ is the flux calculated using the benchmark surface
(the fitting formulas) at the $i$-th phase bins and $N_{\rm bins} = 16$ is total
number of phase bins used.
Second, we define the ``normalized'' residual
\begin{equation}
    \mathscr{M} \equiv \frac{1}{N_{\rm bins}} \frac{\sum_{i=1}^{N_{\rm bins}} |F^{\rm B}_i - F^{\rm fit}_i|}{\langle F^{\rm B} \rangle}\,,
\label{eq:m_def}
\end{equation}
where $\langle F^{\rm B} \rangle$ is the mean value of the benchmark pulse profile,
\begin{equation}
\langle F^{\rm B} \rangle \equiv \frac{1}{N_{\rm bins}} \, \sum_{i=1}^{N_{\rm bins}} F^{\rm B}_i \,.
\end{equation}

\begin{figure*}
\includegraphics[width=2\columnwidth]{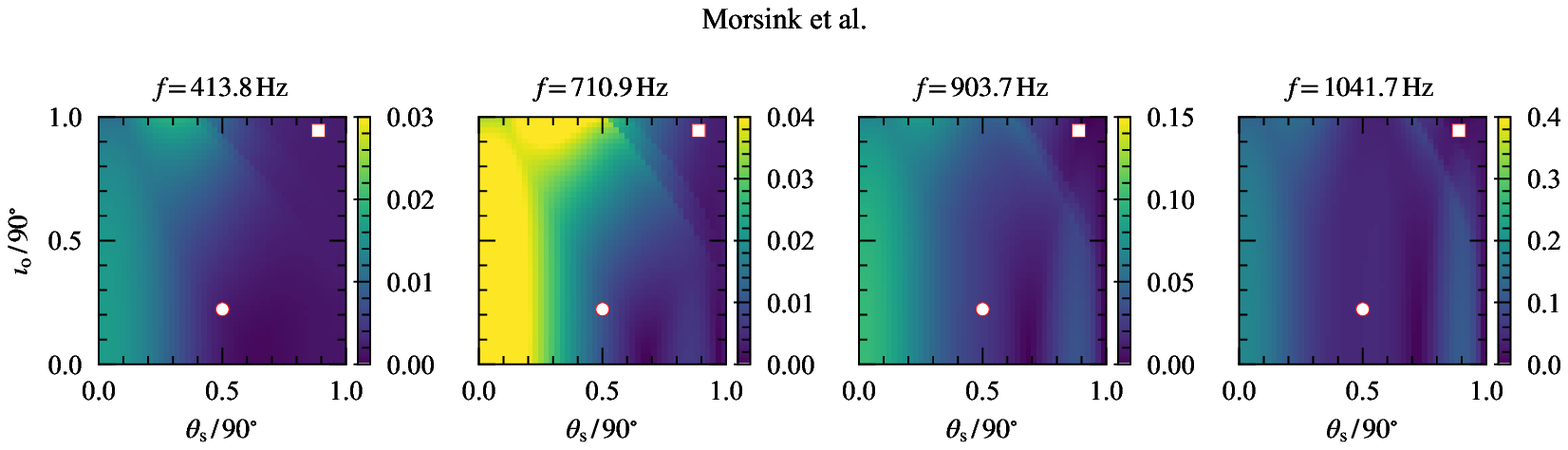} \\
\includegraphics[width=2\columnwidth]{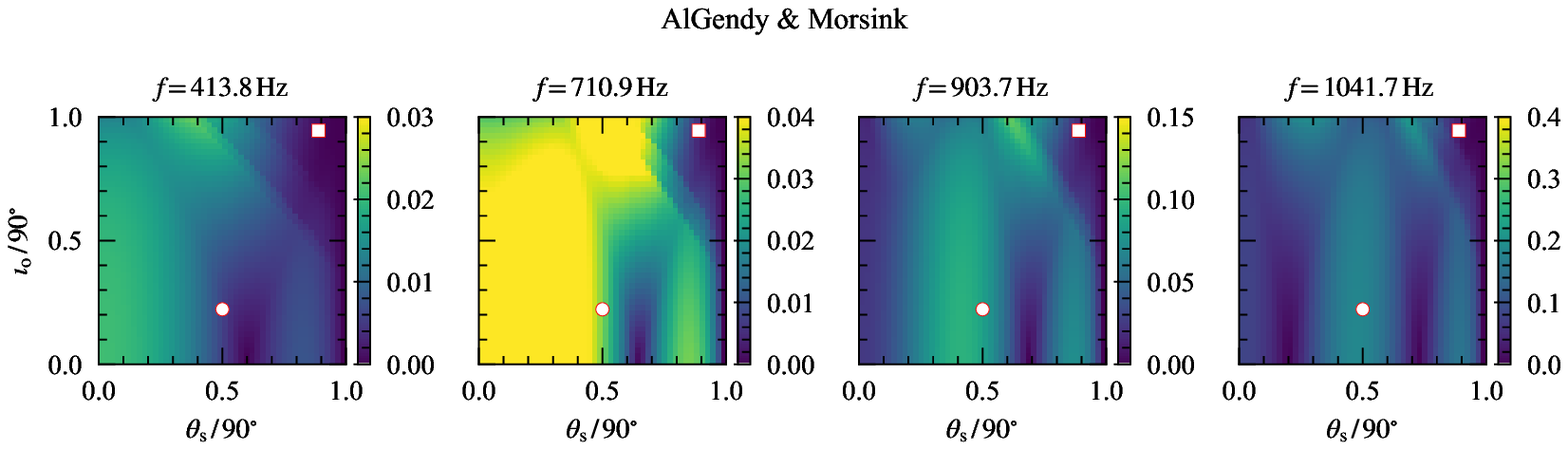} \\
\includegraphics[width=2\columnwidth]{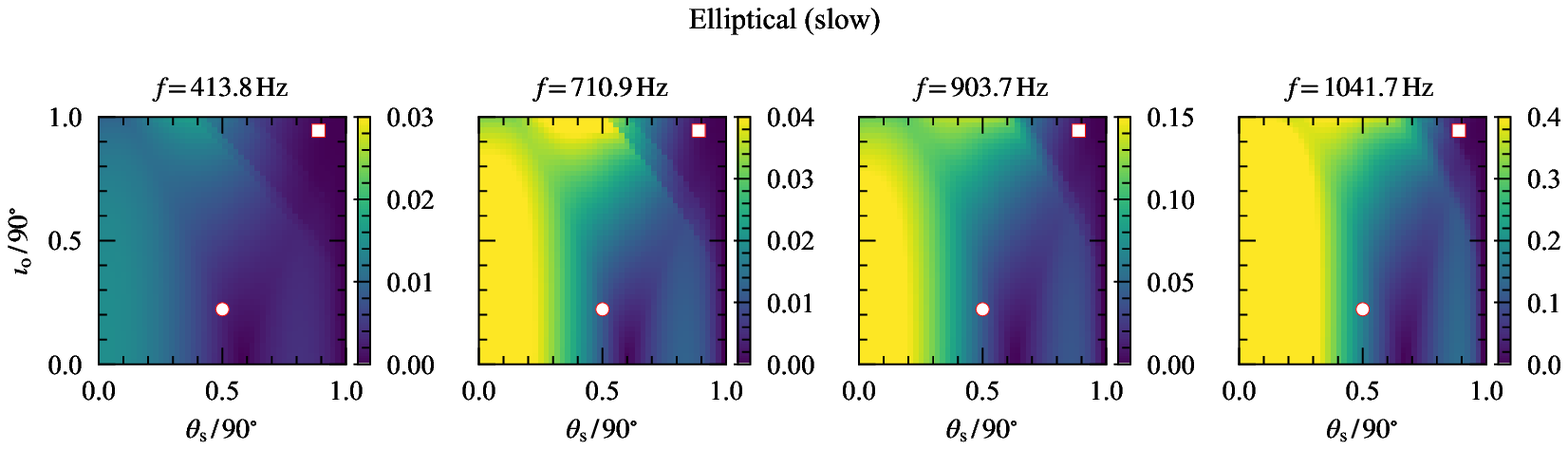} \\
\includegraphics[width=2\columnwidth]{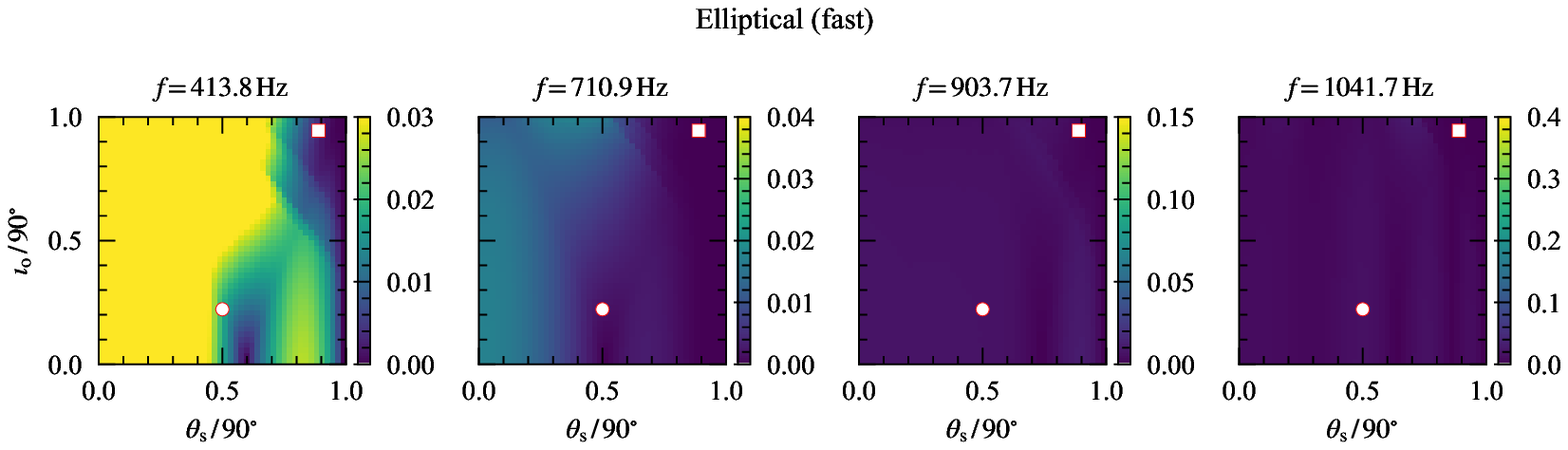}
\caption{The normalized residual $\mathscr{M}$ [defined in
Eq.~\eqref{eq:m_def}] as function of ($\theta_{\rm s}$,~$\iota_{\rm o}$). The
results for $\mathscr{R}$ [defined in Eq.~\eqref{eq:r_def}] are
qualitatively the same. We use the same scale for all panels
in each column. The columns correspond to which benchmark star we compared each fitting formula against.
We see that as the rotation frequency $f$ increases the \ms (top row) and \ag
(middle row) in general deteriorate relative to our benchmark fluxes,
calculated using the formulas of Sec.~\ref{sec:fit_comparison}.
On the other hand, the elliptic fit remains relatively accurate for the whole
frequency range considered by us.
This conclusion can be quantified by calculating the integrals of $\mathscr{M}$ and $\mathscr{R}$, whose results are shown in Fig.~\ref{fig:cumulative_errors_versus_freq}.
The markers denote the two combinations of hot spot colatitude ($\theta_{\rm s}$) and line of sight to the observer ($\iota_{\rm o}$) angles used in our
parameter estimation study in Sec.~\ref{sec:var_fit_formulas}.}
\label{fig:error_angle_dependence}
\end{figure*}

In Fig.~\ref{fig:error_angle_dependence}, we show  $\mathscr{M}$ as a function of
($\theta_{\rm s}, \iota_{\rm o}$) in the domain $\mathscr{D} = [0, 90^{\circ}] \times [0, 90^{\circ}]$,
for four sample benchmark stars with rotation frequencies $413.8$,~$710.9$,~$903.7$~and~$1041.7$~Hz.
These correspond to the stars labeled 1, 3, 4 and 7 in Table~\ref{tab:ref_stars}.
These four stars define the columns in Fig.~\ref{fig:error_angle_dependence},
while the four fitting formulas define the rows.
We use the same color map scale along each column.
This figure reveals a number of interesting facts, namely:
\begin{itemize}
\item As expected, the normalized residual is minimal at $\theta_{\rm s} = \iota_{\rm o} = 90^{\circ}$.
In fact, it remains small for any $\iota_{\rm o}$, as long as $\theta_{\rm s} \approx 90^{\circ}$,
for all formulas.
\item The \ms, \ag and (slow) elliptical formulas all have small
normalized residuals for all combinations of $\theta_{\rm s}$ and $\iota_{\rm o}$
relative to the benchmark flux at $413.6$~Hz (leftmost column).
Since this value is already larger than the fastest
spinning neutron star in NICER's target list, we can expect that these three
formulas would imply similar best fit parameter estimates if used to
analyze NICER data.
Perhaps unsurprisingly, the fast-elliptical fit (which was obtained using
only $\sigma \geq 0.25$ stars) has regions in the $(\ts,\,\io)$ where the normalized
residual becomes larger ($\mathscr{M} \geq 0.3$).
Yet, these regions are confined to $\ts \leq 25^{\circ}$.
\item For faster rotating stars (the three rightmost columns), we see that the
\ms, \ag and slow elliptical formulas start to fail to reproduce the benchmark
flux, as can be seen by the increase in size of the region in which
$\mathscr{M} \gtrsim 0.3$.
There are regions however, where the normalized residual still remains small.
In contrast, the fast elliptical formula outperforms all the three formulas when
applied to rapidly rotating stars, as we should expect, by construction.
\end{itemize}

\begin{figure}
\includegraphics[width=\columnwidth]{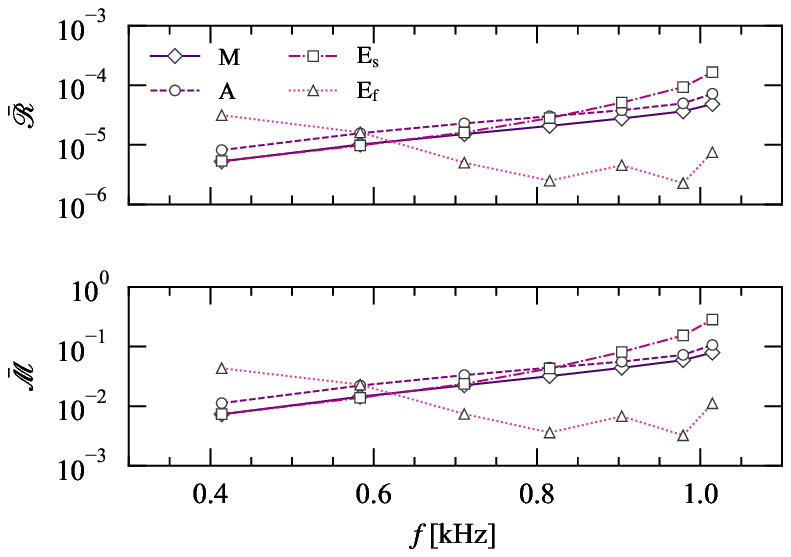}
\caption{The integrated values of $\bar{\mathscr{R}}$ (top panel) and $\bar{\mathscr{M}}$ (bottom panel) in the
domain shown in Fig.~\ref{fig:error_angle_dependence} as functions of the rotation frequency $f$
of the benchmark stars. The legend is shared between both panels.
Overall, all three formulas perform well for slowly rotating stars. As the
rotation frequency $f$ increases, the errors for the \ms (solid lines) and \ag (dashed lines)
fits increase, whereas for the elliptic fit (dash-dotted line) the errors show less variability.}
\label{fig:cumulative_errors_versus_freq}
\end{figure}

In Fig.~\ref{fig:cumulative_errors_versus_freq} we show the dimensionless integrated values
of $\mathscr{M}$, defined as
\begin{equation}
\bar{\mathscr{M}} \equiv \int_{\mathscr{D}} \mathscr{M}
\, \dd (\theta_{\rm s} / 90^{\circ})
\, \dd (\iota_{\rm o} / 90^{\circ})\,,
\end{equation}
(and likewise for $\bar{\mathscr{R}}$)
as a function of the rotation frequency $f$
for the seven benchmark stars.
The figure shows that these two error measures behave similarly. For the \ms, \ag and slow-elliptical
fits, both $\bar{\mathscr{R}}$ and $\bar{\mathscr{M}}$ increase monotonically as a function of $f$.
On the other hand, for the fast-elliptical fit both $\bar{\mathscr{R}}$ and
$\bar{\mathscr{M}}$ decrease with $f$ to values smaller than the other three
curves, yet showing a small oscillatory behavior past 700~Hz, probably
associated with numerical error.

\subsection{Systematics errors on the equatorial radius inference}
\label{sec:parameter_estimation}

In this section we study how the different formulas used to describe surface of
rotating neutron stars affect the parameter estimation of the star's
equatorial radius. We continue to use the simplifying assumptions of
Sec.~\ref{sec:lightcurves} and the parameters summarized in Table~\ref{tab:lightcurve_params}.
We further fix the orientation angles $(\theta_{\rm s}$, $\iota_{\rm o})$
according to the two cases listed in Table~\ref{tab:cases}.
Finally, the star parameters $M$, $R_{\rm eq}$ and $f$ are fixed to:
\begin{itemize}
\item $M=1.4$~M$_{\odot}$, $R_{\rm eq} = 13$~km and $f=205$~Hz as to mimic
the parameters inferred from PSR~J0030+0451~\cite{Miller:2019cac}. We use the \ag
formula to describe the star's surface. (Section~\ref{sec:var_fit_formulas}.)
\item $\{M,\, R_{\rm eq},\, f\}$ are those of the benchmark stars of Table~\ref{tab:ref_stars}.
We use the benchmark surface models of Sec.~\ref{sec:fit_comparison}.
\end{itemize}

In both cases, we use the same methodology to perform a
(restricted) likelihood analysis study as used in~\cite{Silva:2019leq}.

\subsubsection{Statistical methods}
\label{sec:methods}
We call the signal measured during an observation the
{\it synthetic injected signal}, or (for brevity)
{\it the injection} $F_{\rm inj} (\bm{\vartheta}^{\ast})$.
The pulse profile that we use to extract and characterize this observed pulse
profile is referred to as {\it the model} $F_{\rm mod}(\bm{\vartheta})$.
Here, ${\bm{\vartheta}^{\ast}}$ (${\bm{\vartheta}}$) represent the injection (model)
parameters used to calculate the pulse profile.

Both pulse profiles are calculated using the O+S approximation once all
parameters
\begin{equation}
\bm{\vartheta} = \{M, R_{\rm eq}, f, \theta_{\rm s}, \iota_{\rm o}, \Delta \theta_{\rm s}, d, k_{\rm B}T'_0\}\,,
\end{equation}
have been specified.
As done in the previous section, we work with a reduced model parameter space
obtained by fixing
\begin{equation}
\bm{\vartheta}_{\rm fix} = \{M, f, \theta_{\rm s}, \iota_{\rm o}, \Delta \theta_{\rm s}, d, k_{\rm B}T'_0 \}\,,
\end{equation}
to the injected values, leaving as the single variable parameter the equatorial radius $R_{\rm eq}$.

We calculate the best-fit parameter value by minimizing the
reduced chi-squared $\chi^2_{\rm red}$ between the injection and the model pulse profiles,
sampling over the model's variable parameter $R_{\rm eq}$.
The reduced chi-squared is defined as
\begin{equation}
\chi_{\rm red}^2 \equiv
\frac{1}{N}\sum_{i = 1}^{N}
\left[
\frac{F_{\rm mod}(\phi_i, {\bm \vartheta}_{\rm fix}, R_{\rm eq}) - F_{\rm inj}(\phi_i, {\bm \vartheta}_{\rm fix}, R_{\rm eq}^{\ast})}{\sigma(\phi_i)}
\right]^2\,, \\
\label{eq:red_chi}
\end{equation}
where $R^{\ast}_{\rm eq}$ is the equatorial radius of the star used to
calculate the injection pulse profile.
The summation in~\eqref{eq:red_chi} is over the $N$ time stamps during the course
of one observed revolution of the star.
We normalize the phase (dividing by $2\pi$) for a revolution such that $\phi_i \in [0, 1]$
and use $N = 16$ time stamps.
The standard deviation of the distribution ($\sigma$) is modeled as
%
% \begin{equation}
$\sigma(\phi_i) =
\sigma_{R_{\rm eq}}(\phi_i)$,
% \end{equation}
%
where $\sigma_{R_{\rm eq}}$ is the standard deviations on the (injection) equatorial radius.
We calculate the standard deviation $\sigma_{R_{\rm eq}}$
as~\cite{Ayzenberg:2016ynm,Ayzenberg:2017ufk}.
\begin{align}
\sigma_{R_{\rm eq}} &= \frac{1}{2}
\left\vert
F_{\rm inj}(\phi_i, {\bm \vartheta}_{\rm fix}, R_{\rm eq}^{\ast} + \delta R_{\rm eq}) \right.
\nonumber \\
&\quad \left.
    - \,\, F_{\rm inj}(\phi_i, {\bm \vartheta}_{\rm fix}, R_{\rm eq}^{\ast} - \delta R_{\rm eq})
\right\vert\,,
\end{align}
where we assume the values for the statistical error $\delta R_{\rm eq}$
listed in Table~\ref{tab:cases}.
To obtain the standard deviation, we need to calculate the pulse profile
emitted by a star with radii $R^{\ast}_{\rm eq} \pm \delta R_{\rm eq}$.
In this calculation, we cannot use the ``exact'' fits (because they are valid \emph{only} for
the benchmark stars), nor the fitting formulas we are using to calculate the model
pulse profile $F_{\rm mod}$ (because it could bias the resulting likelihood).
To overcome this problem, we obtained a high-order \ag fit, similar to
Eq.~\eqref{eq:fit_algendy_etal} but adding terms $a_{2i} \, \mu^{2i}$ up
to $i=5$ and using \emph{only stars described by the SLy4
equation of state.}

Once the reduced chi-squared is obtained, we assume that the likelihood is Gaussian
\begin{equation}
L(R_{\rm eq}) = \exp\left(- \chi_{\rm red}^2 / 2\right)\,,
\end{equation}
which we combine with the prior $\pi(R_{\rm eq})$, to obtain the posterior
\begin{equation}
P(R_{\rm eq}) \propto L(R_{\rm eq}) \cdot \pi(R_{\rm eq})\,.
\label{eq:posterior}
\end{equation}
We use a flat prior in the range $\kappa \in [0.125, 0.3125]$ for the
compactness~\cite{Miller:2019cac}.
We also set an upper bound on the spin parameter, $\sigma \leq 1$, a
condition that is only violated by stars rotating near their mass-shedding
frequency.
These two conditions combined with the fixed mass $M$ and rotational frequency
$f$ of the star (used to produce the injection pulse profile) fix a range of
values for $R_{\rm eq}$.
We take our prior on $R_{\rm eq}$ to be uniform in the range $R_{\rm eq} \in [10,\, R_{\rm eq}^{\rm max}]$~km,
where the upper bound is set by the lower and upper limits on $\kappa$ and $\sigma$
respectively.

\begin{figure}
\includegraphics[width=\columnwidth]{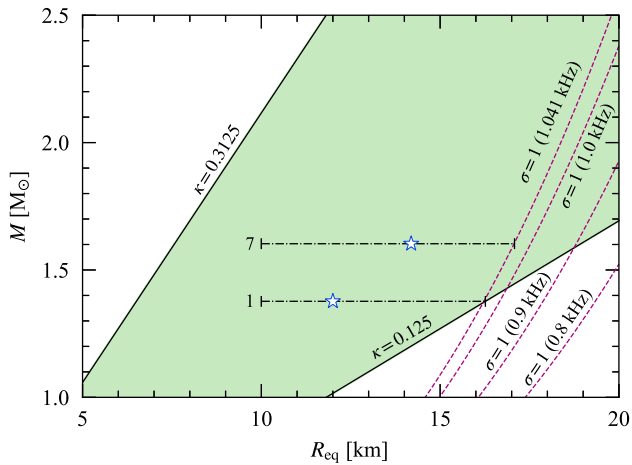}
\caption{Lines of constant compactness ($\kappa$) and spin parameter ($\sigma$)
in the mass-equatorial-radius plane. The two solid lines correspond to
the edges of the prior range on the compactness.
The four dashed lines mark curves of constant $\sigma = 1$ for some sample
rotational frequency values: $f=$~800,~900,~1000 and 1041~Hz.
For stars with $M \lesssim 1.4$~M$_{\odot}$, the lower end of the prior
($\kappa = 0.125$), fixes the largest allowed value of the equatorial radius
$R^{\rm max}_{\rm eq}$.
For stars with larger masses, the upper limit $\sigma \leq 1$,
reduces $R^{\rm max}_{\rm eq}$ if the rotation frequency $f$ is
sufficiently high.
Both scenarios are illustrated by the dot-dashed lines labeled 1 and 7,
which make reference to the labels used in Table~\ref{tab:ref_stars}.
For the line labeled 1, the mass is $1.377$~M$_{\odot}$ and $f = 413.8$~Hz and
thus $R_{\rm eq}^{\rm max}$ is set by lower end of the prior in the compactness.
Conversely, for the line labeled 7, the mass is $1.603$~M$_{\odot}$ and $f=1041.7$~Hz
and thus $R_{\rm eq}^{\rm max}$ is set by the $\sigma = 1$ ($f=1041.7$~Hz) curve.
The values of $R_{\rm eq}$ and $M$ of the benchmark stars 1 and 7 are marked with stars.
}
\label{fig:const_kappa_sigma}
\end{figure}

In Fig.~\ref{fig:const_kappa_sigma} we illustrate this discussion.
The solid lines delimit the allowed region in the ($M$,\,$R_{\rm eq}$)-plane by
the compactness prior alone.
Part of this region is carved out by imposing an upper limit on $\sigma$ which,
for four sample values of $f$, are shown by the dashed lines.
We see that for the slowest rotating star in Table~\ref{tab:ref_stars} (for
which $1.377$~M$_{\odot}$ and $f = 413.8$~Hz) the value of $R_{\rm eq}^{\rm
max}$ is set by the lower prior on the compactness ($\kappa = 0.125$).
On the other extreme, for the fastest rotating star (for which
$1.603$~M$_{\odot}$ and $f=1041.7$~Hz), the value of $R_{\rm eq}^{\rm max}$ is
set by the upper bound on the spin parameter ($\sigma = 1$ with $f \approx
1041$~Hz).
The prior ranges on $R_{\rm eq}$ for these two examples are illustrated by the
dot-dashed lines labeled ``1'' and ``7'', respectively.

To obtain the posterior distribution $P(R_{\rm eq})$, we evaluate
Eq.~\eqref{eq:posterior} on a fine grid covering
$R_{\rm eq} \in [10,\,R^{\rm max}_{\rm eq}]$~km.
Next, we sort the pair $\{R_{{\rm eq}, i},\, P(R_{{\rm eq}, i})\}$ in an
descending order of posterior. The first entry determines the best fit inferred
value of the equatorial radius.
We are also interested in the $1\sigma$ credible intervals of the resulting
posterior distributions.
To calculate them, we add all $P(R_{{\rm eq}, i})$-values until the cumulative
sum reaches 68\% of the total $\sum_{i}^{N} P(R_{{\rm eq}, i})$.
The smallest and largest values of $R_{{\rm eq}, i}$ in this interval yield the
credible interval.

\subsubsection{Systematics due to fitting formulas}
\label{sec:var_fit_formulas}

In this section, we calculate our injection flux using the \ag model for the
star surface, assuming $M^{\ast} = 1.4$~M$_{\odot}$,  $R_{\rm eq}^{\ast} =
13$~km and $f^{\ast} = 205$~Hz. These values were chosen to mimic a source similar to
PSR~J0030+0451 as inferred by the Illinois-Maryland analysis~\cite{Miller:2019cac}.
We are interested in whether the other formulas (\ms and
elliptical) can recover the injected equatorial radius.

\begin{figure}
\includegraphics[width=\columnwidth]{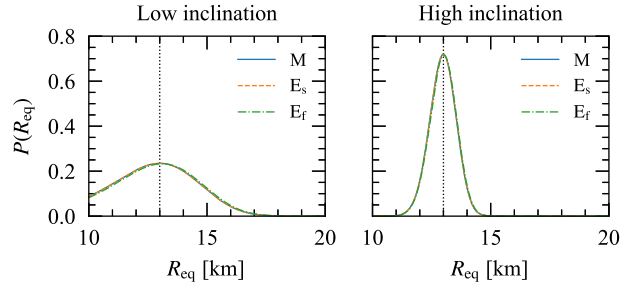}
\caption{Posterior probability distributions of the equatorial radius. Left
panel: for the low-inclination ($\theta_{\rm s} = 45^{\circ}$, $\iota_{\rm o} =
20^{\circ}$) orientation.  Right panel: for the high-inclination ($\theta_{\rm s}
= 85^{\circ}$, $\iota_{\rm o} = 80^{\circ}$) orientation.
Pulse profile models using both the \ms (solid lines) and the
elliptic fits (dashed and dot-dashed lines) recover the injected radius $R_{\rm eq}^{\ast} =
13$~km (vertical dotted line).
}
\label{fig:radius_posterior_algendy_ref}
\end{figure}

In Fig.~\ref{fig:radius_posterior_algendy_ref} we show the resulting posterior
distributions on $R_{\rm eq}$ obtained from this exercise, which we did for the two
hot spot-observer orientations of Table~\ref{tab:cases}.
The posteriors clearly show that the best fit values of $R_{\rm eq}$ for both
\ms (solid lines) and the two flavors of the elliptical formulas
(dot-dashed and dashed lines) agree well with the injection $R_{\rm eq}^{\ast}$
(vertical dotted line).

These results are hardly surprising given our discussion in
Sec.~\ref{sec:dependence_orientation} but serve (albeit through a restrictive
likelihood analysis) to show that all three formulas work equally well in
describing the pulse profile emitted by neutron stars targeted by NICER, i.e.
millisecond pulsars with rotation frequencies below a few hundred
hertz~\cite{Bogdanov:2019ixe}.

\subsubsection{Systematics due to equation of state averaging}
\label{sec:var_eos}

We now turn our attention to the systematic errors that may be introduced by the fact that
the surface formulas represent an average of the shape of an ensemble of
neutron stars, described by different equations of state and spanning
various frequencies, while the target is described by a single equation of state.
To do this, we use the stars from Table~\ref{tab:ref_stars} to calculate the injection
pulse profiles with their surfaces modeled using the formulas described in Sec.~\ref{sec:fit_comparison}.
Next, we perform the same likelihood analysis described in
Sec.~\ref{sec:methods}, using in our model each of the surface formulas, and
then, we analyze the resulting posterior distributions.
These steps are repeated for both hot spot-observer orientations listed in Table~\ref{tab:cases}.

\begin{figure}
\includegraphics[width=\columnwidth]{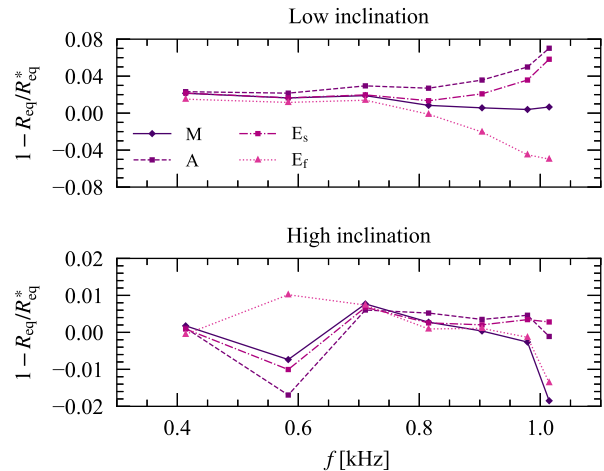}
\caption{Fractional difference between injected $R_{\rm eq}^{\ast}$ and inferred best fit equatorial
radii $R_{\rm eq}$ as a function of the rotation frequency $f$ for the different fitting formulas.
For the low-inclination orientation the error stays below 10\%, with the
``fast'' elliptic fit recovering the injected $R_{\rm eq}$ at high frequencies.
For the high-inclination orientation all formulas recover the injected $R_{\rm eq}$
with less than 2\% precision independently of $f$.
In both cases panels, all fractional errors are smaller than the statistical
errors we have assumed (0.2 for the low inclination and 0.05 for the high inclination scenarios).
}
\label{fig:error_best_fit}
\end{figure}

Figure~\ref{fig:error_best_fit} summarizes our findings and constitutes the main
results of this paper. The figure shows the fractional error between the best
fit value of the equatorial radius (as inferred by a given surface formula) and
the injected value of the equatorial radius as a function of the rotation
frequency.
Different markers correspond to the different fitting formulas.
In the top panel (corresponding to the low-inclination orientation) we see that
all formulas recover well the injected equatorial radius $R^{\ast}_{\rm eq}$ at small
frequencies, with the fast elliptical formula working surprisingly well in this situation.
As we increase the rotation frequency $f$, we see that the \ag and slow elliptic
formula increasingly underestimate $R^{\ast}_{\rm eq}$, in the worst scenario by 7\%
and 6\% percent respectively.
A similar behavior is seen for the fast elliptical fit, which tends to
overestimate $R^{\ast}_{\rm eq}$ instead, but by a similar percentage.  In
contrast, the \ms fit inference remains robust over the whole $f$ range,
misinferring the equatorial radius by $\sim 3\%$ at most (for the fastest
spinning star).
In the bottom panel (corresponding to the high-inclination orientation), we see that
all formulas recover accurately $R^{\ast}_{\rm eq}$ regardless of the spin
frequency of the star, with errors staying below 2\%.

What are the implications of these results to real data analysis with NICER?
Bearing in mind the oversimplifications we have used in our data analysis study, our results
indicate that the systematic error introduced by the averaging procedure in
obtaining the fitting formulas used to model the pulse profile emission of neutron stars
is subdominant relative to the statistical error, which in our case is modeled
by the value of $\delta {R_{\rm eq}}$, that is, below 20\% for the low-inclination orientation
and 5\% for the high-inclination orientation.
In Table~\ref{tab:recoverd_vals} we show the median and the $\pm 1 \sigma$ interval
for the inferred equatorial radii using the various fitting formulas.

An interesting result of our calculation is that the \ag formula, despite its
simple form, is sufficient to infer the injected radii $R^{\ast}_{\rm eq}$ with
percent fractional difference smaller than 6\%, even for the fastest rotating star.
Is this because we used rapidly rotating models when obtaining our own version
of the \ag~fit?
To answer this question, we repeated our analysis, but using the same
coefficients $c_{(i,j)}$ from Ref.~\cite{AlGendy:2014eua} (quoted between
parenthesis in Table~\ref{tab:coefs_ma}).
As we mentioned before, the original \ag fit used only slowly rotating stars
with spin parameter $\sigma < 0.1$.
The outcome of this result is surprising: \emph{the percent fractional difference
remains a few percent, even in the extreme case of the fastest rotating star}.
Quantitatively, in the low inclination orientation the percent fractional difference
increase in magnitude from 5.8\% to 7.6\% for the fastest rotating star.
In the high inclination orientation, this value \emph{decreases} from 0.5\% to 0.1\%.
(See Table~\ref{tab:recoverd_vals_original}, which also includes the results of
the same exercise, but using the \ms~fit~\cite{Morsink:2007tv}.)
The conclusion is then clear: we have found evidence that the original \ag
formula~\cite{AlGendy:2014eua} has a domain of applicability wider than
originally expected.

\begin{table*}[t]
\begin{tabular}{c c c c c c c}
\hline
\hline
Model & $R^{\ast}_{\rm eq}$ & $f$       & $R_{\rm eq}^{\rm M}$ & $R_{\rm eq}^{\rm A}$ &  $R_{\rm eq}^{{\rm E}_{\rm s}}$ & $R_{\rm eq}^{{\rm E}_{\rm f}}$ \\
      & (km)                & (Hz)      & (km)                 & (km)                 & (km)          & (km)      \\
\hline
1     & 12.00               & 413.8
& ($11.74^{+1.73}_{-1.48}$) $11.98^{+0.50}_{-0.49}$
& ($11.72^{+1.72}_{-1.46}$) $11.99^{+0.50}_{-0.49}$
& ($11.74^{+1.73}_{-1.47}$) $11.99^{+0.50}_{-0.49}$
& ($11.82^{+1.73}_{-1.47}$) $12.01^{+0.51}_{-0.50}$
\\ \hline
2     & 12.27               & 583.4
& ($12.07^{+1.58}_{-1.39}$) $12.36^{+0.36}_{-0.40}$
& ($12.00^{+1.55}_{-1.34}$) $12.48^{+0.38}_{-0.38}$
& ($12.07^{+1.58}_{-1.39}$) $12.39^{+0.37}_{-0.39}$
& ($12.13^{+1.57}_{-1.38}$) $12.15^{+0.35}_{-0.40}$
\\ \hline
3     & 12.57               & 710.9
& ($12.42^{+1.43}_{-1.31}$) $12.56^{+0.49}_{-0.48}$
& ($12.29^{+1.37}_{-1.22}$) $12.58^{+0.50}_{-0.48}$
& ($12.41^{+1.43}_{-1.29}$) $12.57^{+0.48}_{-0.49}$
& ($12.48^{+1.42}_{-1.32}$) $12.56^{+0.72}_{-0.49}$
\\ \hline
4     & 12.90               & 815.4
& ($12.79^{+1.28}_{-1.25}$) $12.86^{+0.45}_{-0.47}$
& ($12.55^{+1.17}_{-1.08}$) $12.83^{+0.47}_{-0.48}$
& ($12.73^{+1.25}_{-1.16}$) $12.87^{+0.45}_{-0.46}$
& ($12.92^{+1.30}_{-1.37}$) $12.89^{+0.44}_{-0.45}$
\\ \hline
5     & 13.27               & 903.7
& ($13.20^{+1.13}_{-1.22}$) $12.27^{+0.52}_{-0.53}$
& ($12.80^{+0.98}_{-0.92}$) $13.23^{+0.51}_{-0.51}$
& ($13.00^{+1.05}_{-0.97}$) $13.25^{+0.51}_{-0.51}$
& ($13.54^{+1.30}_{-1.66}$) $13.26^{+0.52}_{-0.50}$
\\ \hline
6     & 13.70               & 978.9
& ($13.64^{+1.02}_{-1.24}$) $13.73^{+0.50}_{-0.53}$
& ($13.01^{+0.79}_{-0.76}$) $13.63^{+0.49}_{-0.52}$
& ($13.21^{+0.83}_{-0.76}$) $13.65^{+0.48}_{-0.50}$
& ($14.31^{+1.46}_{-1.77}$) $13.72^{+0.44}_{-0.48}$
\\ \hline
7     & 14.19               & 1041.7
& ($14.09^{+0.94}_{-1.22}$) $14.45^{+0.37}_{-0.62}$
& ($13.20^{+0.60}_{-0.59}$) $14.20^{+0.37}_{-0.55}$
& ($13.36^{+0.61}_{-0.55}$) $14.15^{+0.30}_{-0.51}$
& ($14.89^{+1.47}_{-1.24}$) $14.38^{+0.21}_{-0.45}$
\\ \hline
\hline
\end{tabular}
\caption{
Inferred equatorial radii $R_{\rm eq}$ for each fitting formula.
The first three columns correspond to the benchmark star label, its equatorial radius
and its rotation frequency, respectively.
The remaining columns are the median and the $\pm 1\sigma$ credible intervals as
inferred by using the different fitting formulas in the pulse profile model.
For the inferred $R_{\rm eq}$ entries, the results between parenthesis correspond to the
low-orientation case, while the others to the high-inclination case.}
\label{tab:recoverd_vals}
\end{table*}

\begin{table*}[t]
\begin{tabular}{c c c c c}
\hline
\hline
Model & $R^{\ast}_{\rm eq}$ & $f$       & $R_{\rm eq}^{\rm M}$ as in~\cite{Morsink:2007tv} & $R_{\rm eq}^{\rm A}$ as in~\cite{AlGendy:2014eua}  \\
      & (km)                & (Hz)      & (km)                                             & (km)                                               \\
\hline
1     & 12.00               & 413.8
& ($11.72^{+1.72}_{-1.47}$) $11.96^{+0.49}_{-0.48}$
& ($11.74^{+1.73}_{-1.47}$) $11.99^{+0.50}_{-0.49}$
\\ \hline
2     & 12.27               & 583.4
& ($12.02^{+1.56}_{-1.37}$) $12.44^{+0.38}_{-0.39}$
& ($12.04^{+1.56}_{-1.37}$) $12.38^{+0.37}_{-0.39}$
\\ \hline
3     & 12.57               & 710.9
& ($12.34^{+1.40}_{-1.30}$) $12.55^{+0.50}_{-0.48}$
& ($12.35^{+1.39}_{-1.25}$) $12.57^{+0.48}_{-0.49}$
\\ \hline
4     & 12.90               & 815.4
& ($12.67^{+1.25}_{-1.21}$) $12.82^{+0.47}_{-0.49}$
& ($12.65^{+1.21}_{-1.13}$) $12.86^{+0.45}_{-0.47}$
\\ \hline
5     & 13.27               & 903.7
& ($13.02^{+1.10}_{-1.15}$) $12.24^{+0.53}_{-0.54}$
& ($12.94^{+1.04}_{-0.98}$) $13.24^{+0.52}_{-0.50}$
\\ \hline
6     & 13.70               & 978.9
& ($13.38^{+0.96}_{-1.11}$) $13.72^{+0.52}_{-0.57}$
& ($13.20^{+0.84}_{-0.84}$) $13.66^{+0.50}_{-0.51}$
\\ \hline
7     & 14.19               & 1041.7
& ($13.73^{+0.84}_{-1.06}$) $14.45^{+0.47}_{-0.64}$
& ($13.42^{+0.66}_{-0.66}$) $14.26^{+0.36}_{-0.56}$
\\ \hline
\hline
\end{tabular}
\caption{
Inferred equatorial radii $R_{\rm eq}$ using the formulas of \ms~\cite{Morsink:2007tv}
and \ag~\cite{AlGendy:2014eua}, and the fitting
coefficients obtained originally in these works.
The table is analogous to Table~\ref{tab:recoverd_vals}.
Overall, the relative difference ($\equiv 2 \cdot |a-b|/|a+b|$) between
the median values of $R_{\rm eq}$ using our and the original fitting
coefficients is below 3\%.
Among both formulas, the largest fractional difference between best-fit
$R_{\rm eq}$ and the injected $R_{\rm eq}^{\ast}$ values happens for the \ag formula (7.6\%,
Model 7).
}
\label{tab:recoverd_vals_original}
\end{table*}

\section{Conclusions}
\label{sec:conclusions}

We studied the systematic error introduced by the use of analytical formulas to
describe the surface of rapidly rotating neutron stars.
These formulas are constructed by fitting certain analytical expressions to an
ensemble of neutron star models described by a variety of equations of state
and covering a wide range of compactness and spin parameter values.
Neutron stars, however, are believed to be described by a single equation of
state, and therefore, the fitting procedure used to obtain these surface formulas
introduces a source of systematic error in the parameter estimation of
neutron star properties, which could have implications to x-ray
pulse profile observations with NICER.

To study the impact of this systematic error, we performed a restricted
likelihood analysis using synthetic pulse profile data.
We found that the systematic error described above is smaller than the
statistical error indicating, albeit in a simplified analysis, that the
radius parameter estimation by NICER~\cite{Riley:2019yda,Miller:2019cac} is not
affected by it.
It would be interesting to repeat the analysis carried here in a complete
set-up following, for instance, the theoretical studies
in~\cite{Lo:2013ava,Lo:2018hes,Miller:2016kae}, using as the injection pulse
profile (i.e.~synthetic signal) one calculated using the ``exact'' formulas
obtained here.
More specifically, it would be interesting to investigate the \emph{cumulative
effect} of this systematic error when one considers multiple finite-sized hot
spots~\cite{Sotani:2019cqv,Sotani:2020ipl} and how it depends on their location
on the star's surface.
As seen in Fig.~\ref{fig:error_angle_dependence} this error has a nontrivial
behavior in the case of a single, pointlike hot spot.
It would be important to analyze it in more realistic hot spot geometries
ideally reproducing the hot spot configurations as inferred by NICER for
PSR~J0030+0451~\cite{Riley:2019yda,Miller:2019cac}.
We think it is unlikely that this systematic error will matter for the
slowly-spinning neutron stars targeted by NICER, but we hope our work motivates
further studies, which should also include the level of realism of
a full statistical analysis as done in~\cite{Riley:2019yda,Miller:2019cac}.

Another interesting question to explore is how our ignorance on the equation
of state affects the resulting fitting formulas.
In our analysis, we used for our synthetic data the pulse profile emitted from
the surface of a neutron star whose equation of state was \emph{also} used to
obtained the fitting formulas.
In practice, it is unlikely that this situation will happen and
it would then be important to investigate the variability (and the
implications to radii inferences) of using different equation of state catalogs
which could differ from the one used here to produce the surface fits.

Finally, it would also be important to repeat this analysis in the context of
future large-area x-ray timing facilities~\cite{Watts:2018iom}, such as the \emph{enhanced X-ray
Timing and Polarimetry} (eXTP)~\cite{Zhang:2018edu} and the \emph{Spectroscopic
Time-Resolving Observatory for Broadband Energy X-rays}
(STROBE-X)~\cite{Ray:2018dlb,Ray:2019pxr} missions.
These future missions are expected to provide more precise parameter estimation
of the radii of neutron stars relative to NICER's current capabilities.
As the statistical error is decreased, all sources of systematic errors will
become more important, and the one discussed here may be of relevance.

As by-products of our study we also presented a method to accurately locate
the surface of rotating neutron star solutions obtained with RNS.
An implementation of the method is publicly available in~\cite{GPRNSRepo}.
Moreover, we have also introduced a new analytical formula to describe the
surface of rapidly rotating neutron stars. This formula, based on the ellipsoidal
isodensity approximation~\cite{Lai:1993ve}, better captures the surface of
rapidly rotating neutron stars relative to other formulas known in the
literature.
The application range of this new formula is not limited by the problems
studied here, and it could also be used to model the effect of stellar
oblateness on parameter estimation using the cooling tail
method~\cite{Suleimanov:2020ijb} or in the wave propagation on thin oceans on
neutron star surfaces~\cite{vanBaal:2020imd}.

\section*{Acknowledgments}
It is a pleasure to thank Fred Lamb, Cole Miller, Stuart Shapiro, Hajime
Sotani, Nikolaos Stergioulas, and Helvi Witek for various discussions.
We are particularly indebted to Sharon Morsink for helping us implement our O+S
code, sharing with us numerical data used to validate it and numerous
discussions on the subject.
We also thank the anonymous referee for carefully reading our work and
the suggestions to improve it.
H.O.S. and G.P. thank Hajime Sotani and the Division of Theoretical Astronomy of
the National Astronomical Observatory of Japan (NAOJ) through the NAOJ Visiting
Joint Research supported by the Research Coordination Committee, NAOJ, National
Institutes of Natural Sciences (NINS) for the hospitality while part of this
work was done.
N.Y. thanks the hospitality of the Kavli Institute for Theoretical Physics
(KITP) while part of this work was carried.
The work of H.O.S. and N.Y. was supported by the NSF Grant No.~PHY-1607130 and
NASA grants No.~NNX16AB98G and No.~80NSSC17M0041.
G.P. acknowledges financial support provided under the European Union's H2020
ERC, Starting Grant agreement no. DarkGRA-757480.
K.Y. acknowledges support from NSF Award No.~PHY-1806776, NASA Grant No.~80NSSC20K0523,
a Sloan Foundation Research Fellowship and the Owens Family Foundation.
K.Y. would like to also acknowledge support by the COST Action GWverse~CA16104
and JSPS KAKENHI Grants No.~JP17H06358.

\appendix

\section{``Exact'' surfaces: numerical derivatives, error estimates, and fits}
\label{app:exact}

In this appendix we show the details behind the fits for the star surface $R$ and
its logarithmic-derivative $\dd \log R / \dd \theta$ used to model the shape
of our benchmark stars.

First, to assess the numerical error associated with the surface data $R$ we
computed neutron star solutions with two different resolutions using the RNS
code.

The RNS code solves for the neutron star model's interior and exterior
on a grid with the radial coordinate $r$ compactified and equally spaced
in the interval $s \in [0,1]$, using the definition $s \equiv r/(r+r_{\rm eq})$,
and the angular coordinate $\mu = \cos\theta$ also equally spaced
in the interval $\mu \in [0,1]$.
This way, the code assigns half of the grid to the interior of the star (the
equatorial location of the surface is always at $s=1/2$) and the other half to
the vacuum exterior. The radial resolution near the surface, if we assume that
we have chosen $S$ grid points, will be
\begin{equation}
\Delta r|_{r_{\rm eq}}\sim r_{\rm eq}  \left(\frac{4}{2+S}\right)\,,
\end{equation}
which for a star with $r_{\rm eq}$ approximately 10~km and grid sizes of
$S=301$ and $S=1201$ points is around $0.13$ and $0.033$ km respectively.
The usual choice for the angular grid is to be half of the radial one.
Therefore in our calculations we have used both a low resolution grid
of size $301\times 151$ points and a high resolution grid of size
$1201\times 601$ points.

Once a neutron star solution is obtained, with either resolution, the
star's surface is obtained by the loci of the circumferential radius
where the enthalpy per unit mass becomes equal to zero [see Eq.~\eqref{eq:circ_radius}].

To obtain an estimate on the numerical error on $R$ for our high-resolution
solution ($\epsilon_{\rm high}$), we calculate the maximum fractional difference between the
high-resolution and low-resolution solutions (evaluated at the same grid points
$\mu_i$), namely
\begin{equation}
\epsilon_{\rm high} = \textrm{max} \vert 1 - R_{\rm low}(\mu_i) / R_{\rm high}(\mu_i) \vert\,.
\label{eq:error_estimate}
\end{equation}
We find that $\epsilon_{\rm high}$ is of the order of $10^{-5}$ for all stars in Table~\ref{tab:ref_stars}.
We used the high-resolution data to obtain all fits.

\begin{figure*}[t]
\includegraphics[width=\columnwidth]{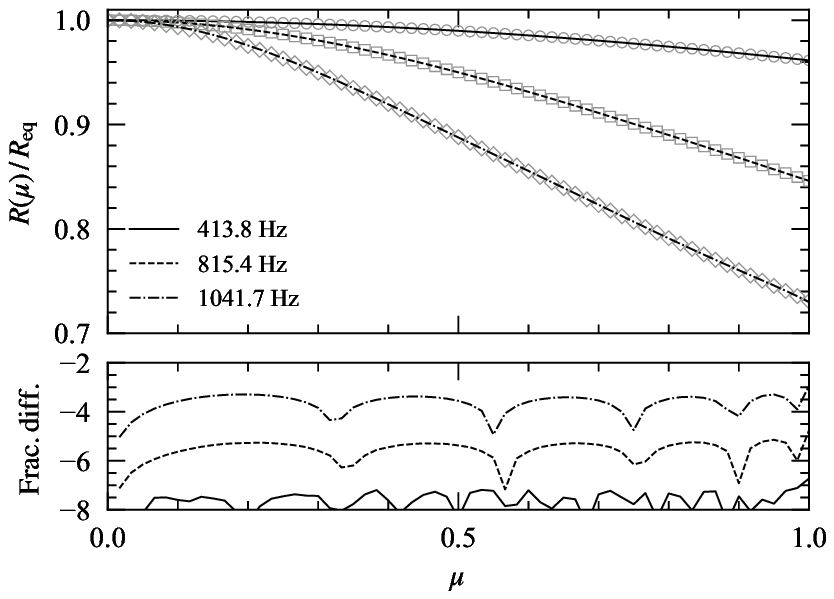}
\includegraphics[width=\columnwidth]{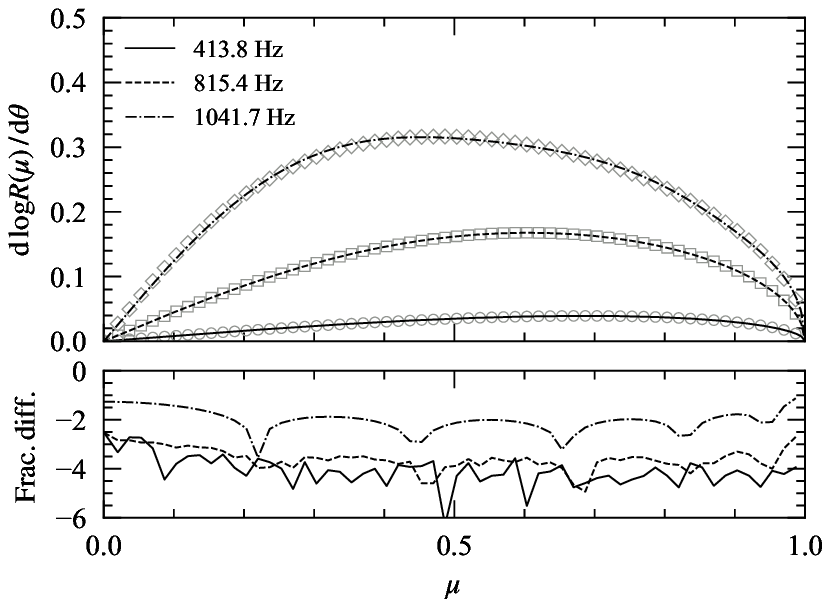}
\caption{Surface and logarithmic-derivative of the three sample
benchmark star models (with spin frequencies $f = 413.8$, $815.4$, and~$1041.7$~Hz)
as functions of the colatitude $\theta$.
In all panels, the markers correspond to the numerical data, whereas the lines
to the fitting formulas~\eqref{eq:exact_fits}.
Left-top panel: the star's surface normalized relative to its equatorial radius
as a function of $\mu = \cos\theta$.
Right-top panel: the star's logarithmic-derivative relative to $\theta$ also as
a function of $\mu$.
Bottom panels: the fractional differences
$\log_{10} |1 - y_{\rm fit} / y_{\rm data}|$
between fit ($y_{\rm fit}$) and numerical data ($y_{\rm data}$).
}
\label{fig:exact_surfaces}
\end{figure*}

\begin{figure*}
\includegraphics[width=\columnwidth]{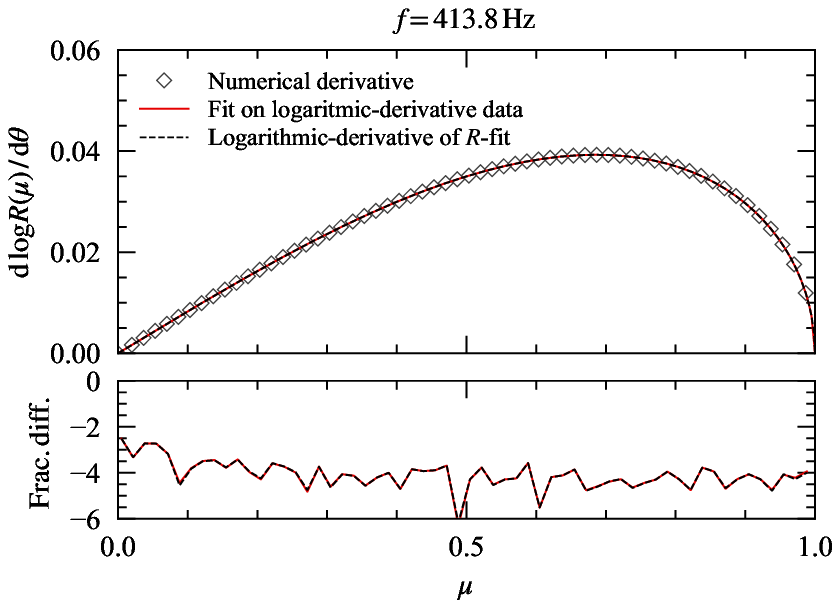}
\includegraphics[width=\columnwidth]{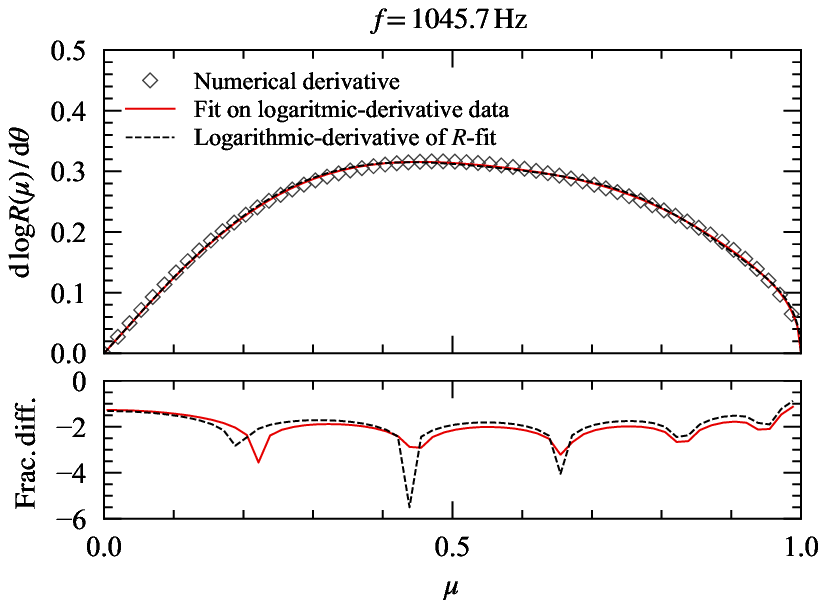}
\caption{The logarithmic-derivative of the surface data. In the top panels the markers show the
logarithmic derivative calculated using a sixth-order central finite difference scheme [Eq.~\eqref{eq:fd}].
The dashed lines show the fits directly applied to this data using Eq.~\eqref{eq:dlogr_exact_fit} and the solid
lines show the predicted logarithmic-derivative obtained by first applying Eq.~\eqref{eq:r_exact_fit} to $R(\mu)$ and
\emph{then} taking the logarithmic-derivative.
The fractional differences
$\log_{10} |1 - y_{\rm fit} / y_{\rm data}|$
between fit ($y_{\rm fit}$) and numerical data ($y_{\rm data}$)
are shown in the bottom panels.
The left panels corresponds to a star with spin frequency $f = 413.8$~Hz,
whereas the right panels corresponds to a star with spin frequency
$f = 1045.7$~Hz.
For the slowest spinning case (left figure), both approaches agree well with the
numerical data. However, for the fastest spinning case (right figure) the
fractional differences are, except at a few points, slightly larger when using
Eq.~\eqref{eq:r_exact_fit}. This fact justifies the use of a separate fit based
on Eq.~\eqref{eq:dlogr_exact_fit} to model the surficial numerical derivatives.
}
\label{fig:exact_dlogr}
\end{figure*}

In Fig.~\ref{fig:exact_surfaces} we show the surface (left panel) and its
logarithmic-derivative (right panel) as functions of $\mu = \cos\theta$ for
three sample benchmark stars.
We find that our fits (curves) agree with the numerical data (markers) within
less then 1\% fractional differences for all three cases.

In Fig.~\ref{fig:exact_dlogr} we show the logarithmic derivative of the stellar surface as function of $\mu$ for the slowest (left panel) and
the fastest (right panel) spinning benchmark star. In both panels the markers show the logarithmic derivative obtained by applying Eq.~\eqref{eq:fd}
to the high-resolution numerical data.
The curves correspond to two approaches to model this data.
More specifically, the solid curves correspond to fits obtained by directly
applying Eq.~\eqref{eq:dlogr_exact_fit} to fit the data, while the dashed
curves correspond to first fitting $R(\mu)$ using Eq.~\eqref{eq:r_exact_fit}
and then taking the logarithmic derivative.
We see that both approaches agree very well for the $f = 413.8$ Hz spinning star. However, for the $f = 1045.7$ Hz spinning star,
the former approach performs better overall, except at a few points. As a consequence, we used a separate fit based on Eq.~\eqref{eq:dlogr_exact_fit}
to model the surficial numerical derivatives.

Let us now consider the logarithmic-derivative of $R$, defined in Eq.~\eqref{eq:log_deriv_definition}
\begin{equation}
\frac{\dd \log R(\mu)}{\dd \theta} = - (1-\mu^2)^{1/2} \frac{1}{R(\mu)} \frac{\dd R(\mu)}{\dd\mu}\,.
\nonumber
\end{equation}
We calculate the derivative numerically using our high-resolution surface data and
using a six-order central finite difference formula,
\begin{align}
\frac{\dd R}{\dd \mu} &= \left[ R(\mu + 3 \Delta \mu)
- 9 R(\mu + 2 \Delta \mu)
+ 45 R(\mu + \Delta \mu)
\right.
\nonumber \\
&\quad \left.
- 45 R(\mu - \Delta \mu)
+ 9  R(\mu - 2 \Delta \mu)
-    R(\mu - 3 \Delta \mu)
\right]
\nonumber \\
&\quad \cdot (60 \, \Delta \mu)^{-1} + {\cal O}(\Delta\mu^6) \,,
\label{eq:fd}
\end{align}
where $\Delta \mu$ (the $\mu$-grid size) is approximately $1.67 \cdot 10^{-3}$.

We quantify the error on the numerical derivative by doing the
calculation at two different resolutions $\Delta \mu$ and $2 \Delta \mu$.
Using Eq.~\eqref{eq:error_estimate}, we find that the error using the
finer grid varies between approximately $5 \cdot 10^{-4}$ for the
fastest rotating star and $7 \cdot 10^{-3}$ for the slowest rotating star.

\clearpage
\bibliography{biblio}

\end{document}